\documentclass[11pt]{article}

\usepackage{graphicx}
\usepackage{graphicx,psfrag}

\def \v {\vskip 0.2cm }

\def \s {\sigma}

\def \i {{\hbox {i}}}

\def \l {\lambda}

\def \cd {{\cal D}}

\def \ce {{\cal E}}

\def \cn {{\cal N}}

\def \cv {{\cal V }}

\def \U {\Upsilon} 

\def \vep {\varepsilon}

\def \a {\alpha}

\def \b {\beta}

\def \T {{\hbox {\ Tr\,}}}

\def \th {\theta}

\def \S {{\cal S}}

\def \cc {{\cal C}}

\def \cu {{\cal U}}

\def \L {\Lambda}

\def \ce {{\cal E}}

\def \cy {{\cal Y}}

\def \ca {{\cal A}}

\def \g {\gamma}

\def \G {\Gamma}

\def \h {\hat}

\def \E {{\bf E}}

\def \i {{\hbox{ i}}}

\def \vep {\varepsilon}

\def \th {\theta}

\def \D {\Delta}

\def \d {\delta}

\begin{document}
\title{On Connected Diagrams  and Cumulants of Erd\H os-R\' enyi Matrix Models}

\author{O. Khorunzhiy\thanks{Universit\'e de Versailles -- Saint-Quentin, Versailles 78035 FRANCE; e-mail: \mbox{khorunzhiy@math.uvsq.fr,} Tel.: +33 1 39 25 36 25}}

\date{}

\maketitle

\begin{abstract}
Regarding the adjacency matrices of $n$-vertex graphs and related graph Laplacian, 
we introduce  two families of discrete matrix models constructed both with the help of 
the Erd\H os-R\'enyi ensemble of random graphs. Corresponding matrix sums 
represent the characteristic functions of the average number of  walks and 
closed walks over the random graph. These sums can be considered as  discrete analogs of the matrix integrals of 
random matrix theory.

We study the diagram structure of the cumulant expansions of logarithms 
of these matrix sums and  analyze  the limiting expressions as $n\to\infty$
in the cases of constant and vanishing edge probabilities.

\end{abstract}


\section{Introduction}

It counts about thirty years from the time when the map enumeration problems of theoretical physics were
related with the diagram representation of the formal cumulant expansions of the
form
$$
{1\over N^2} \log  \E  \left\{ e^{g V_N} \right\} = \sum_{k\ge 1} {g^k\over k! }\, {1\over N^2} Cum_k(V_N), 
\eqno (1.1)
$$
where the average is taken with respect to the gaussian measure over the space of hermitian $N$-dimensional 
matrices $\{H_N\}$ and $V_N$   is given by $\T H_N^q$, or more generally,  by a linear combination of such traces with certain degrees  $q\ge 3$ .
This relation between maps, diagrams and random matrices observed first by t'Hooft 
is still a source of numerous results 
that reveal deep links between
various branches  of mathematics and theoretical physics (see e.g. papers 
\cite{BIZ,DGZ,O}  for the references and  reviews of results).  
In these studies,  the nature of the large-$N$ limit  of the right-hand side of (1.1) and existence of its asymptotic expansion in degrees of $1/N^2$
are the problems of the primary importance \cite{BI,EM,Gu}. 

In present paper we examine the diagram structure of a discrete analog of (1.1),  where the role of $H_N$ is played by the $n\times n$ 
adjacency matrix $A_n$  of the Erd\H os-R\'enyi random graphs with the edge probability $p_n$. 
In this setting,  $V_N$ of (1.1) is replaced by random variables
$$
X_n^{(q)} = \T A_n^q = \sum_{i=1}^n (A^q_n)_{ii}
\quad
{\hbox{and}} \quad  
Y_n^{(q)} = \sum_{i,j=1}^n (A_n^q)_{ij},
\quad q\ge 2
\eqno (1.2)
$$
and this gives two different 
 discrete matrix models related with the powers of adjacency matrix and of the Laplace operator on graphs.
Variables $X$ and $Y$ represent the numbers of   close walks of $q$ steps and walks of $q$ steps  over the graphs, respectively.

We study the limiting behavior of the cumulants of  $X_n^{(q)}$ and $Y_n^{(q)}$
in the following three asymptotic regimes,  when $p_n = O(1)$, 
${1\over n} \ll p_n\ll 1$, or $p_n = O(1/n)$ as $n\to\infty$. 
Our main result is  that 
in all of these three regimes, 
the following limits exist (cf. (1.1))
$$
{1\over p_n n^2} Cum_k\left(g_nY_n^{(q)}\right) \to F_k^{(q)}(\omega), \quad n\to\infty, 
\eqno (1.3)
$$
with the appropriate choice of the normalizing factors $g_n$.
Limiting expressions depend on asymptotic regime indicated by 
 $\omega\in \{1,2,3\}$.
 The same statement as (1.3) holds for random variables $X^{(q)}_n$ with corresponding changes of $g_n$ and the  limiting expressions.

To prove convergence (1.3), we develop a diagram technique similar to that commonly adopted  in the random matrix theory. 
We show that  the diagram structure of the cumulants of (1.3) is closely related with the trees with  $k$ labeled edges. 
In the simplest case of $q=2$ and $\omega=2$, we derive explicit recurrent relations that determine the limits (1.3) and 
prove that the corresponding exponential generating function verifies the 
Lagrange (or P\'olya) equation. Regarding other values of $q$ and $\omega$, 
we obtain several generalizations of this equation. 

 Using (1.3), we  show that  the Central Limit Theorem 
 is valid for the centered and renormalized variables $X_n^{(q)}$ and $Y_n^{(q)}$. 
 This describes asymptotic properties of the averaged numbers of walks and closed walks over the random graphs. 
Our results imply
CLT for the normalized spectral measure of the adjacency matrices of Erd\H os-R\'enyi 
random graphs. This improves known results about the convergence of the normalized spectral measure \cite{BG,KSV}.
Also, we indicate an asymptotic regime when the CLT does not hold for variables $X_n^{(q)}$ and $Y_n^{(q)}$.

The paper is organized as follows.
In Section 2 we determine two families of matrix sums that can be called 
the Erd\H os-R\'enyi matrix models. 
The key observation here is that the graph Laplacian generates
the  Erd\H os-R\'enyi measure on graphs.
This allows us to identify $Y^{(2)}_n$ as a natural  analog of the
quartic potential $V_N=\T H_N^4$ of (1.1). 
In Section 3, we develop a general diagram technique to study the cumulant expansions 
of the form (1.1) for discrete Erd\H os-R\'enyi matrix models of the first and the second type (1.2). 
In Section 4, we
 prove convergence of normalized  cumulants  in three main asymptotic regimes. 
 The issue of  the Central Limit Theorem  is discussed at the end of Section 4.  
In Section 5, we study the classes of connected diagrams and 
derive recurrent relations for their numbers.
In Section 6, we complete the study of limiting expressions $F_k^{(q)}(\omega)$ (1.3) and 
consider the formal limiting transition 
for the free energy. This free energy takes different forms in dependence on the edge probability as $n\to\infty$.
Section 7 contains a
summary  of our results.

\section{Discrete Erd\H os-R\'enyi  matrix models}

In (1.1), one can rewrite mathematical expectation as follows
$$
 \E  \left\{ \exp \left( {g\over N} \T H^q_N\right) \right\} =
{1\over C_N} \int _{ {\cal H}_N} \exp\left\{ - \beta \T H^2_N + {g\over N} \T H^q_N\right\}dH_N,\ \beta >0
 \eqno (2.1)
$$
where $H_N$ is $N\times N$ hermitian matrix, $C_N= C_N(\beta)$ is a normalization constant, and  the matrix integral runs over the space ${\cal H}_N$ of all hermitian matrices 
with respect to the Lebesgue measure $dH_N$. The ensemble of random matrices $\{H_N\}$ distributed according 
to the gaussian measure with the density $\exp\{-\b \T H_N^2\}$ is known as the Gaussian Unitary invariant
Ensemble abbreviated as GUE . This ensemble plays a fundamental  role in the random matrix theory
(see monograph \cite{Me} and references therein). The matrix integral of (2.1) is  
known as the partition function 
of the matrix model with the potential $V_N = \T H_N^q$.

Regarding $\T H_N^2\ $ of (2.1) as a formal kinetic energy term, one can speculate about the 
trace of the Laplace operator; being restricted to the space of functions on graphs, it
leads immediately to a discrete analog of the matrix integral of (2.1). In this case
the integral $\int_{{\cal H}_N}$ of (2.1)  is replaced by the sum over the set of all $n$-dimensional adjacency
matrices of graphs.
This approach is  fairly natural  and one benefits from two important counterparts of it.

From one hand, the use of graph Laplacian 
indicates a natural analog of  the matrix models with quartic potentials and clarify
 relations between the weights generated by $\T H^2$ and $\T H^4$.  This helps to distinguish 
two matrix models, the traditional one given by  $X_n^{(q)}$ and the new one determined by $Y_n^{(q)}$ (1.2).

From another hand, a simple but important reasoning relates the  graph Laplacian
with the Erd\H os-R\'enyi probability measure on graphs. 
On this way, 
the  gaussian measure of  GUE in the average of the left-hand side of (2.1) is 
replaced by a measure 
with nice  properties  we are going to describe.

\subsection{Graph Laplacian and Erd\H os-R\'enyi random graphs}

Given  a finite graph with the set  of $n$  labeled vertices ${\cal V}_n = \{v_1,\dots , v_n\}$ and 
 the set  of simple non-oriented edges $E_m= \{e^{(1)}, \dots , e^{(m)}\}$, the discrete analog of the 
 Laplace operator $\D(\g)$ on the graph $\g$  is defined  by relation
$$
\D(\gamma) = \partial ^* \partial,
\eqno (2.2)
$$
where $\partial $ is the difference operator determined on the space of complex functions on vertices $\cv_n\to {\bf C}$ and $\partial^*$ is its conjugate determined on the space of complex functions on edges $E_m\to {\bf C}$ (see  \cite{Mo} for more details).

It can be easily shown that in the canonical basis, the linear  operator 
$\D(\gamma)=\D_n $ has 
$n\times n$ matrix
with the elements 
$$
\D_{ij} = \cases{ \deg(v_j) ,& if $i=j$, \cr
-1, & if $i\neq j$ and $(v_i,v_j)\in E$,\cr
0, & otherwise,\cr
}
\eqno (2.3)
$$
where $\deg(v)$ is the vertex degree. 
If one considers the $n\times n$ adjacency matrix $A = A(\g)$ of the graph $\g$, 
$$
A_{ij} = \cases{ 1 , & if $(v_i,v_j)\in E$, $i\neq j$,\cr
0, & otherwise,\cr}
$$
then one can rewrite the definition of  $\D$ (2.2) in the form 
$$
\D_{ij} = B_{ij} - A_{ij} \quad {\hbox {with}} \ \ B_{ij} = \delta_{i,j} \sum_{l=1}^n A_{il}, 
\eqno (2.4)
$$
where $\delta_{i,j}$ is the Kronecker $\d$-symbol
$$
\delta_{i,j} = \cases{1, & if $i=j$, \cr
0, & if $i\neq j$.}
$$
It follows from (2.2) that $\D(\gamma_n)$ has positive eigenvalues.

Let us consider the set $\G_n$ of all possible simple non-oriented graphs $\g_n$ with the set $\cv=\cv_n $ of $n$ labeled vertices. Obviously, $\vert \Gamma_n\vert = 2^{n(n-1)/2}$.
Given an element $\gamma\in \G_n$, it is natural to consider the trace $\T \D(\gamma)$ as 
the  "kinetic energy" of the graph $\gamma$. Then we can 
assign to each graph $\gamma_n$ the Gibbs weight
$\exp\{ - \b \T \D(\g_n) \}$, $\beta >0$ and  introduce the discrete analog of the integral (2.1) by relation 
$$
Z_n(\b,Q) = \sum_{\g_n \in \G_n } \exp\{ - \b  \T \D_n + Q(\g_n)\},
\eqno (2.5)
$$
where $\D_n = \D(\gamma_n)$ and $Q$ is an application: $\Gamma_n \to {\bf R}$ that we specify later. 

Let us note that  one should normalize the sum  (2.5)  by 
$\vert \G_n\vert$, 
but 
this does not play any role with respect to further results. In what follows, we omit subscript $n$ in $\D_n$.
Relation (2.4) implies that
$$
\T \D = \sum_{i=1}^n \D_{ii} = \sum_{i,j=1}^n A_{ij} = 2\sum_{1\le i< j\le n} A_{ij}.
\eqno (2.6)
$$
Then we can rewrite (2.5) in the form
$$
Z_n(\b,Q) = 
\sum_{\g_n \in \G_n } e^{  Q(\g_n)} \prod_{1\le i<j\le n} e^{-2\beta A_{ij}}.
\eqno (2.7) 
$$
It is easy to see that 
$$
Z_n(\b,0) =  \left( 1 + e^{-2\b}\right)^{n(n-1)/2}.
\eqno (2.8)
$$
Then the normalized partition function can be represented as
$$
\h Z_n(\b,Q) = Z_n(\b,Q)/Z_n(\b,0) = \E_{\b} \left\{ e^{Q(\gamma)}\right\},
\eqno (2.9)
$$
where $\E_{\b}\{\cdot \}$ denotes the mathematical expectation
with respect to the measure  supported on the set $\G_n$.
This measure assigns to each element $\gamma \in \G_n$  probability
$$
P_n(\gamma) = {\displaystyle e^{-2\beta \vert E(\gamma)\vert}\over 
 \left( 1 + e^{-2\b}\right)^{n(n-1)/2}},
 $$
 where $E(\gamma) $ denotes the set of  edges of  the graph $\gamma$.

Given a couple $(i,j)$, $i,j \in \{1,\dots,n\}$, one can determine a random variable
$a_{ij}$ 
on the probability space $(\G_n,P_n)$ that is the indicator function of the edge $(v_i,v_j)$
$$
a_{ij} (\gamma)= \cases{ 1, & if $(v_i,v_j)\in E(\gamma)$,\cr 
0, & otherwise.\cr}
$$
It is easy to show that   the random variables $\{ a_{ij}\}_{ 1\le i\le  j\le n}$ are jointly independent and are of the same Bernoulli distribution depending on $\beta$ such that
$$
a_{ij}^{(\b)} =\cases{ 1-\d_{ij} ,& with probability $ {\displaystyle e^{-2\beta}\over 
\displaystyle 1+ e^{-2\beta}}=p$,
\cr
0, & with probability $1-p$.\cr }
\eqno (2.10)
$$
The term $1-\d_{ij}$ reflects the property that graphs $\gamma$ have no loops.

The probability space $(\G_n, P_n)$ is known as the Erd\H os-R\'enyi (or Bernoulli) ensemble of random graphs with the edge probability $p$ \cite{JLR}. Since the series of pioneering
papers by Erd\H os and R\'enyi, 
the asymptotic properties of graphs  $(\G_n, P_n)$, 
such as the size and the number of 
connected components, the maximal and minimal vertex degree and many others, are extensively studied (see \cite{B,JLR}).
Spectral properties of corresponding random matrices $A$ (2.3) and $\D$ 
(2.4) are considered in a series of  papers (in particular, see \cite{BG,KV,KS,KKM,KSV,MF,RB}).
In present paper we  study the random graph ensemble 
$(\G_n, P_n)$ from another point of view motivated by the 
asymptotic behavior of partition functions (2.9).

\subsection{Quartic potential and Erd\H os-R\'enyi matrix models}

Let us determine the discrete analog of the integral (2.1) with quartic potential $\T H_N^4$. 
Once $\T H^2$ replaced by $\T (\partial^*\partial) = \T \D$ (2.2), 
it is natural to consider 
$$
\T (\partial^*\partial \partial^*\partial) = \T \D^2
$$ 
as the analog of $\T H^4$. Then the partition function 
(2.5) reads as
$$
Z_n(\beta, g) =
\sum_{\g_n \in \G_n } \exp\{ - \b \  \T \D_n + g_n\T \D^2\},
\eqno (2.11)
$$
where $g_n$ is to be specified. It follows from (2.3) and (2.4) that 
$$
\T \D^2 = \T B^2 + \T A^2= \sum_{i,j=1}^n (A^2)_{ij} + \sum_{i,j=1}^n A_{ij}.
$$
Then, using (2.6) and repeating computations of (2.7) and (2.8), we obtain representation
$$
\h Z_n(\b, g_n )= Z_n(\b, g_n )/ Z_n(\b, 0 ) = \left( { 1 + e^{-2\b'}\over 
1+ e^{-2\b}} \right)^{n(n-1)/2} \ \E_{\b'} \{ e^{g_nY_n}\},
\eqno (2.12) 
$$
In this relation, we have 
denoted $\b' = \b - g_n$ and introduced variable
$$
Y_n = \sum_{i,j,l=1}^n a_{il} a_{lj},
\eqno (2.13)
$$
where  $a_{ij}$ are jointly independent random variables of  the law (2.10) with 
$\beta $ replaced by $\beta'$. 
The average $\E_{\b'}$ denotes the corresponding mathematical expectation.
In what follows, we omit the subscripts $\beta$ and $\beta'$
when they are not necessary.

On can generalize  (2.11) and  consider 
mathematical expectation
$$
\E \left\{ \exp ( g_n Y^{(q)}_n)\right\}, \quad {\hbox{with}\quad } 
Y^{(q)}_n = \sum_{i,j=1}^n (A^q)_{ij},
\eqno (2.14)
$$
as the normalized partition function of the discrete  Erd\H os-R\'enyi  matrix model that we will refer to as 
$q$-step walks model.
Also one can consider the average 
$$
\E \left\{ \exp ( g_n X^{(q)}_n)\right\}, \quad {\hbox{with}\quad } 
X^{(q)}_n = \T (A^q),
\eqno (2.15)
$$
that we relate with the discrete Erd\H os-R\'enyi model for $q$-step closed walks. 
These models are different analogs of the matrix integral (2.1).

Finally, let us point out  one more analogy between the  discrete model (2.11) and 
the gaussian matrix integrals (2.1). We mean the invariance property of the 
probability measure with respect to a group of the space transformations \cite{Me}. 

Let $\ca_n$ denotes the set of of all $n$-dimensional symmetric matrices
whose elements are equal to $0$ or $1$ and the diagonal elements are zeros.
It is not hard to see that  if $n\times n$ orthogonal matrix $\Upsilon$ is such that 
$\U A \U^{-1} \in \ca_n$ for all $A\in \ca_n$, then $\U$ verifies the following properties

a) all elements of $\U$ take values $0$ or $1$;

b) any given line of $\U$ contains only one non-zero element and

c) any given column of $\U$ contains only one non-zero element.

\noindent Clearly, the  set of all such orthogonal matrices 
$ \cy_n$ is in one-to-one correspondence with the symmetric group of permutations ${\cal S}_n$.

It is easy to see that  $\U\in \cy_n$ determines a basis change in ${\bf R}^N$ 
because $\U$ re-enumerates 
the vectors of the canonical basis associated with the graph. 
Equality 
$
 \T (\U\D\U^{T}) = \T \D
$
shows  that the probability measure $P$ on $\ca_n$
$$
P(A) = C^{-1} \exp\{ - \b \T \D\}
$$
is invariant with respect the group of transformations $\cy_n$,
$$
P(\U A \U^{T}) = P(A).
\eqno (2.16)
$$

It is not hard to prove the inverse statement: if the probability measure $P$ on the set 
$\ca_n$ verifies (2.16) for all $\U\in \cy_n$, then the random variables given by the matrix elements of $A$ are jointly independent.

This is in complete analogy with the well-known fact of random matrix theory 
that the invariant gaussian distribution  on the space of hermitian (or real symmetric matrices)
generates independent random variables \cite{Me}. 
We do not pursue this topic here and return to asymptotic behavior of  the sums (2.11)
in the limit of infinite $n$.

\section{Cumulant expansions and connected diagrams}

In this section we develop a diagram technique to study the terms of the expansion
$$
\log \E \left\{ e^{gV_n}\right\}  = \sum_{k\ge1} {g^k\over k!}\  Cum_k(V_n),
\eqno (3.1)
$$
where $V_n$ represents  represents $X_n$ (2.15) or $Y_n$ (2.14).
Let us stress that the random variables $a_{ij}$ are bounded and therefore the series (3.1) is absolutely convergent for any finite $n$ and sufficiently small $g$.

Given  a vector $\a = (i_1, \dots, i_r)$, we 
introduce random variables  $U_\a$ 
$$
U_\a = \cases{  
a_{i_1i_2} a_{i_2i_3} \cdots a_{i_qi_{1}}, \ \  r=q 
& for $X$-model \cr
a_{i_1i_2} a_{i_2i_3} \cdots a_{i_qi_{q+1}}, \ \  r=q+1
& for $Y$-model\cr}
\eqno (3.2)
$$
such that 
$V_n = \sum_{ \{\a\}_1^n} U_\a$, where the sum  runs over all possible values of 
$ i_s,\  s=1,\dots, r$.
Then  one can write relation
$$
Cum_k(V_n) = \sum_{\{\a_1\}_1^n}\cdots \sum_{\{\a_k\}_1^n}
 Cum\{U_{\a_1}, \dots, U_{\a_k}\},
\eqno(3.3)
$$
where 
$$
Cum\{U_{\a_1}, \dots, U_{\a_k}\} =  
{d^k\over
dz_1\cdots dz_k} \log \E\{\exp(z_1U_{\alpha_1}+
\dots +z_k U_{\alpha_k})\}\vert_{z_l=0}.
$$
Coefficient  $Cum\{U_{\a_1}, \dots, U_{\a_k}\}= Cum\{ \cu(\vec \a_k)\}, \vec \a_k = (\a_1, \dots, \a_k)$ is known as the semi-invariant of the family 
$\cu(\vec \a_k) = \{ U_{\a_j}\}_{j=1}^ k$ \cite{MM}.

To simplify (3.3), we 
separate the set $\{1, \dots, n\}^{\otimes (rk)}$ into the classes of equivalence according to the 
properties of the family $\cu(\vec \a_k)$. The rule is that given $\vec \a_k$, 
we pay major attention not to the values of variables $i_j$ but rather to the presence of copies of the same random variable $a$
in $\cu(\vec \a_k)$. This approach is fairly common 
in random matrix theory \cite{FK,W}. It leads to the diagram representation 
of the classes of equivalence (see e.g. \cite{DGZ} for the review and for mathematical description).
This method is used to study random matrices with gaussian or centered random variables.  We 
modify it  to the  study of cumulants (3.3) of  the Erd\H os-R\'enyi models.

\subsection{Connected diagrams}

The diagrams we construct for $X$-model and $Y$-model are very similar and we describe 
them in common. 
Let us consider a graph $\l$ with $r$ labeled vertices $\{\th_1, \dots , \th_r\}$ with $r$ determined by (3.2). The graph $\l$ contains $q$ edges:  these are $\vep_j = (\th_j,\th_{j+1})$, $j=1,\dots, q$; 
for the $X$-model we have $\theta_{q+1} = \theta _1$. Given $\a=(i_1, \dots, i_r)$, one can assign 
to each $\theta _j$ the value of $i_j$, $j=1, \dots, r$  and then  each edge $\vep_j$
denotes a  random variable $a_{i_j,i_{j+1}}$.

To study $cum\{\cu\vec (\a_k)\}$ with given $\vec \a_k$, we consider the set 
$\Lambda_k = \{ \l_1, \dots, \l_k\}$ of $k$ labeled graphs $\l$; we will say that $\l_l$ is the 
\textsl{{\bf element}} number $l$ of the diagram we construct and denote by $ \{\vep^{(l)}_j\} _{j=1}^ r$ the edges of this element.
The diagram  consists of $k$ elements and a number of \textsl{{\bf arcs}} $\sigma$ 
that join some of the edges of  $\l$'s. We 
draw an arc $\s_{j,l}^{(j',l')}$ that joins $\vep^{(l)}_j$ and $\vep^{(l')}_{j'}$ when in $\vec \a_k$
$$
i_j^{(l)} = i_{j'}^{(l')}, i_{j+1}^{(l)}=i_{j'+1}^{(l')} \quad {\hbox{or}}\quad 
i_j^{(l)} = i_{j'+1}^{(l')}, i_{j+1}^{(l)}=i_{j'}^{(l')}.
\eqno (3.4)
$$
Here we assume that  $l<l'$. It  follows from (3.4) that the arc can have on of the two  orientations that we call the direct and the inverse one, respectively. If $l=l'$, we assume $j<j'$ and again consider  the 
direct and the inverse arcs.

We say that the edges $\vep_j^{(l)}$ and $\vep_{j'}^{(l')}$ represent  the \textsl{feet} of the arc $\s_{j,l}^{(j',l')}$
 and denote this by
relation $\vep_j^{(l)}\simeq  \vep_{j'}^{(l')}$. Clearly, this relation separates the set of edges 
$$
\ce(k,r) = \{\vep_{j}^{(l)}, j=1, \dots, r; l=1, \dots, k\}
$$ 
 into classes of equivalence that we call the \textsl{{\bf color groups}} of edges. 
It is possible that a class contains one edge only. In this case we  say that there is a 
\textsl{ {\bf simple color group}}. Certainly, we color different classes in different colors.

\begin{figure}[htbp]
\centerline{\includegraphics[width=12cm]{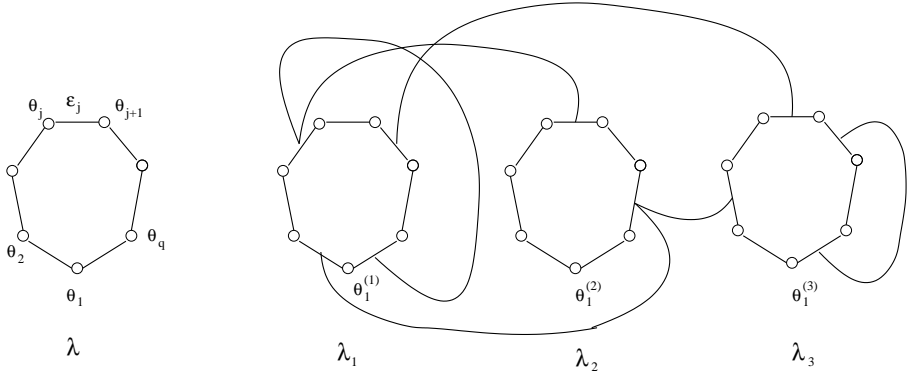}}
\caption{\footnotesize{ A graph $\l$ and a diagram $\d_3$ for $X$-model with $q=7$.}} 
\end{figure}

An important remark is that it is sufficient to consider the\textsl{ {\bf reduced diagrams}}; 
this  means that if there are  
three edges that belong to the same color group,
$\vep_{j}^{(l)} \simeq\vep_{j'}^{(l')}\simeq \vep_{j''}^{(l'')}$, with $l<l'<l''$, then we draw the arcs 
$\s_{j,l}^{(j',l')}$ and $\s_{j',l'}^{(j'',l'')}$
between the nearest neighbors $\vep_{j}^{(l)}, \vep_{j'}^{(l')}$ and $\vep_{j'}^{(l')}, \vep_{j''}^{(l'')}$ only. The same concerns 
the arcs $ \s_{j,l}^{(j',l)}$ whose feet belong both to  $\l_l$. Regarding a reduced diagram, each color group has its \textsl{{\bf minimal edge}} determined in the obvious manner.

To summarize, 
we consider the set $\L_k$ and define  the  application\mbox{ $\vec \a_k \to \d(\vec \a_k)$}
of the set of all $\vec \a_k$ to the set of diagrams
by drawing all arcs prescribed by $\vec \a_k$ 
and reducing of  set of arcs obtained to the set of arcs between nearest neighbors $\Sigma(\vec \a_k)$.
As a result,  we get a   diagram $\d_k = \d (\vec \a_k) = (\L_k, \Sigma(\vec \a_k))$ (see \mbox{Figure 1}).
\v 

We say that  a diagram $\d_k$ is non-connected when there exist at least two disjoint subsets $\tau_1$ and $\tau_2$
of $\{1, \dots, n\}$ such that $\tau_1\cup \tau_2 = \{1, \dots, n\}$ and there is no arc with one foot in 
$\L(\tau_1) = \{ \l_j, j\in \tau_1\}$ and another in $\L(\tau_2)=\{\l_j, j\in \tau_2\}$. If there is no such subsets, then we say that the diagram is connected. The following statement is a well-known fact from the probability theory.

\vskip 0.5cm
{\bf Lemma 3.1} . {\it If $\vec \a_k$ is such that  the corresponding diagram $\d_k = \d_k(\vec \a_k)$ is non-connected, then $Cum\{\cu(\vec \a_k)\}=0$.}
\vskip 0.5 cm
{\it Proof.} It is clear  if there exist two subsets $\tau_1$ and $\tau_2$ as described before, then
the $\s$-algebras generated by   random variables $\{U_{\mu}, \mu  \in \L_1\}$ and 
$\{ U_{\nu}, \nu  \in \L_2\} $ are independent. The fundamental property of semi-invariants is that in this case
$Cum\{\cu(\vec \a_k)\}$ vanishes \cite{MM}.

\subsection{ Cumulants and sums over connected diagrams}

It is clear that for any given set $\vec \a_k = \{ \a_1, \dots, \a_k\}$ there exists only one diagram 
$\d_k = \d(\vec \a_k)$. We agree that two diagrams $\d_k = (\L_k, \Sigma_k)$ and 
$\d_k' = (\L_k', \Sigma_k')$ are not equal, 
 $\d_k\neq \d_k'$ if   in obvious bijection ${\cal J} (\L_k) = \L_k'$ preserving orderings, we have 
  ${\cal J} (\Sigma_k)\neq \Sigma_k'$. 

Let us say that  two vectors $\vec \a_k$ and $\vec  \a'_k$ are equivalent, $\vec \a_k \sim \vec \a'_k$ if
$\d(\vec \a_k) = \d(\vec \a_k')$.  Relation $\sim$ separates the set 
$\{1, \dots, n\}^{\otimes (rk)}$
into the classes of equivalence that we denote by $\cc(\delta_k), \d_k =  \d(\vec \a_k)$. Let  $\cn (\d_k)
=\vert \cc(\delta_k)\vert $ be the cardinality of  the
equivalence class $\cc(\d_k)$.   

Relation (3.4) means that 
two edges $\vep$ and $\vep'$ belong to one color group if and only if  random variables $a$ and $a'$
assigned to these edges by $\vec \a_k$
are equal. This is true for  any given $\vec \a'_k \in \cc(\delta_k)$.
 Let us denote  by $m(\d_k)$ the number of color  groups of $\d_k$. Clearly, 
 random variables that belong to different  groups are jointly independent.

 \vskip 0.5 cm
 {\bf Lemma 3.2.} 
  {\it The right-hand side of relation (3.3) can be represented as 
  $$
 \sum_{\{\a_1\}_1^n}\cdots \sum_{\{\a_k\}_1^n} Cum\{ U_{\a_1}, \dots, U_{\a_k}\}=
\sum_{\d_k\in \cd_k} \cn (\d_k) W(\d_k),
\eqno (3.5)
 $$
 where $\cd_k$ is the set of all possible connected reduced diagrams of the form $(\L_k, \Sigma)$ and
$$
W(\d_k) =  \sum_{\pi_s\in \Pi_k}(-1)^{s-1}(s-1)!\ \left( \E a\right)^{m(\d_k)}
\left( \E a\right)^{\chi(\pi_s,\d_k)}.
\eqno (3.6)
$$
In this relation $\pi_s$ denotes a partition $\pi_s = (\tau_1, \dots \tau_s)$ of the 
set $\{1, 2, \dots, k\}$ into $s$ subsets; $\Pi_k$ is the set of all possible partitions, 
and  $\chi(\pi_s, \d_k)$ 
is the number of additional color groups generated by $\pi_s$.
 }
 
 \vskip 0.3cm 
 {\it Proof.} 
 By definition of the semi-invariant \cite{MM}, we have  that
 $$
 Cum\{U_{\a_1}, \dots, U_{\a_k}\} =  \sum_{\pi_s\in \Pi_k}(-1)^{s}(s-1)!\ 
 \E\{  {\bf U}(\tau_1)\} \cdots \E \{ {\bf  U}(\tau_s)\},
 \eqno (3.7)
 $$
 where we denoted ${\bf U}( \tau_j) =\prod_{\mu\in \tau_j} U_\mu$.
 It is easy to see that   that the right-hand side of (3.7)
  depends on $\d_k$ only, because it  does not change   when  $\vec \a_k$
 is replaced by another  $\vec \a_k'$ from the same equivalence class $\cc(\d_k)$. 
 Denoting by $W(\d_k)$ the right-hand side of (3.7), we get (3.5).

 To prove (3.6), let us choose an element $\vec \a_k^{(0)}$ of the equivalence class 
 $\cc(\d_k)$ and consider  first the trivial partition $\pi_0 = (\tau_0)$ with  $\tau_0= \{1,\dots, k\}$. 
  Then 
 $\chi(\pi_s,\d_k) =0$ and 
 $$
 \E\left\{ V_{\a^{(0)}_1} \cdots V_{\a^{(0)}_k}\right\} = (\E a)^{m(\d_k)},
 $$
 where we have use the independence of random variables $\{a_{ij}\}$ that belong to different color groups of $\d_k$ and the fact that  $a\in \{0,1\}$. 
 
 Now let us consider a partition $\pi_s\neq \pi_0$. It is clear that there exist at least one color group of two or more edges such that its elements belong to two or more different subsets  $\L(\tau_j)= \{ \l_{t},\,  t\in \tau_j\}$. In this case we say that this color group is separated into subgroups; each subgroup contains edges of this color that belong to one subset $T_j$. We color the edges of each new  subgroup in the same color, the edges of the different subgroups - into different colors. 
It follows from (3.7) that 
the color group separated by $\pi_s$ into $v+1$ new subgroups  provides the factor $(\E a)^{v+1}$ to the right-hand side of (3.6). We say that $v$ is the number of additional color groups generated by $\pi_s$.
Regarding all initial color groups, we get factor 
$(\E a)^{m(\d_k)} \, (\E a)^{\Sigma v}$ with $\sum v = \chi(\pi_s,\d_k)$. Lemma 3.2 is proved.

\vskip 0.5cm 
To study $\cn (\d_k)$, let us consider the set of vertices of $\L_k$ 
$$
\Theta_k = \{ \theta_j^{(l)}, l=1, \dots, k;\  j=1, \dots, r\}.
$$
Given $\vec \a_k$, we draw the arcs according to the rule (3.4) and get $\d_k = \d_k(\vec \a_k)$. 
Let us say that equality $i_j^{(l)} = i_{j'}^{(l')}$ of (3.4) identifies corresponding vertices 
$\th_j^{(l)}$ and $\th_{j'}^{(l')}$ and denote this by 
$$
\th_j^{(l)}  \cong \th_{j'}^{(l')}.
\eqno (3.8)
$$ It is easy to see that
relation (3.8) separates $\Theta_k$ into classes of equivalence. 
We determine the minimal element of the class as the vertex whose numbers $j$ and $l$ take minimal values among those of vertices that belong to this class.
If there is no $\th_{j'}^{(l')}$ such that $\th_j^{(l)}  \cong \th_{j'}^{(l')}$, then we say that 
$\th_j^{(l)} $ belongs to the class of equivalence  consisting of one element. Let us denote the total number of the classes of equivalence by $\nu(\d_k)$.

\vskip 0.5cm 
{\bf Lemma 3.3}. {\it Given $\d_k\in \cd_k$ the cardinality $\cn(\d_k) = \vert \cc(\d_k)\vert $ verifies asymptotic relation
$$
\cn (\d_k) = n^{\nu(\d_k)}(1+o(1)),
\eqno (3.9)
$$
 in the limit
$n\to\infty$.
}

\vskip 0.5cm
{\it Proof}. 
It is clear that two equivalent vectors $\vec \a_k\sim \vec \a_k'$ 
generate the same partition of $\Theta_k$ into classes of equivalence. Inversely, given a diagram 
$\d_k$ and regarding corresponding partition of $\Theta_k$, 
we get a separation of the set of variables 
$$
I_k = \{ i_j^{(l)}, \ j=1, \dots, r; \ l=1, \dots, k\}
$$
into $\nu(\d_k)$ groups. Variables that belong to the same group are equal between them. 
To get all possible $\vec \a_k$ from the same equivalence class,
we allow variables $i_j^{(l)}$ to  take all possible values from $1$ to $n$ with obvious restriction that 
variables from different groups take different values. Then obviously
$$
\cn(\d_k) = n(n-1) \cdots (n- \nu(\d_k)+1) = n^{\nu(\d_k)}(1+o(1)), \ \ 
{\hbox{ as }} \ n\to\infty.
\eqno (3.10)
$$
Lemma 3.3 is proved. 
 
 \v 
 Let us complete this subsection with the following useful remark. 
 The set of diagrams $\cd_k$ we constructed gives a graphical representation of vectors
 $\vec \a_k$ and describes the classes of equivalence of such vectors. 
 One can push forward this representation and consider the set of 
 \textsl{graphs of diagrams} $G_k=G(\d_k)$ generated in natural way 
 by diagrams $\d_k$ by gluing the edges of elements $\l$ in the way prescribed by the arcs of $\d_k$. 
 Actually, this is what is usually done in the standard diagram approach of random matrix theory. 

The graph $G(\d_k)= (\hat \Theta, \hat \ce)$ has $\nu(\d_k)$ labeled vertices $\hat \theta$ that correspond
 to the minimal elements of the classes of equivalence of $\Theta_k$. The vertices $\hat \theta $
 and $\hat \theta' $ are joined by an edge $(\hat \theta,\hat \theta')\in \hat \ce$ if there is an edge 
 $\vep\in \ce(k,r)$  that joins corresponding classes of equivalence of $\Theta_k$.  
 Certainly, the number of edges of $G(\d_k)$ is equal to the number of all color groups $m(\d_k)$.

The graph representation is very useful when $\d_k$ contains $k-1$ arcs only. We say that $\d_k$ are tree-like because in this case,
the graphs $G(\d_k)$ of $Y$-model are trees with  color edges.

\begin{figure}[htpb]
\centerline{\includegraphics[width=12cm]{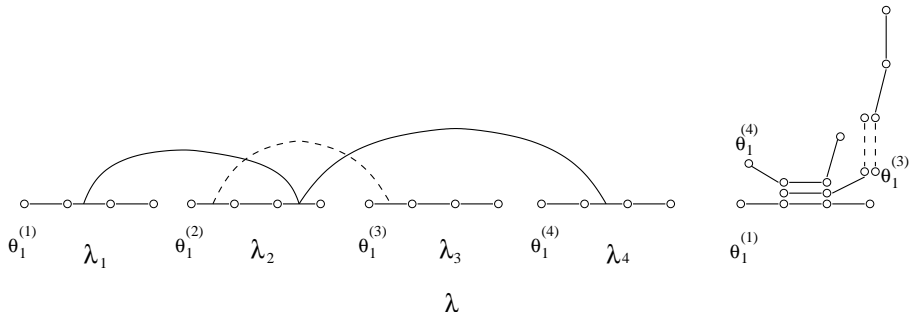}}
\caption{\footnotesize{ Tree-like diagram and corresponding  tree for $Y$-model with $q=3$}}
\end{figure}

Remembering that the arcs $\sigma \in \Sigma$ have the direct and the inverse orientation (3.4),
we agree to consider the graphs $G(\d_k)$ with the  edges of $\d_k$ glued in the inverse sense only.
In what follow, 
we will use representations of $\vec \a_k$ by diagrams $\d_k$ and  graphs $G(\d_k)$ both.
 Note that $^G(\d_k)$ has no loops.

 \section{Diagrams and limits of the cumulants }

 In this  section we characterize the classes of connected diagrams that 
 provide the leading contribution to the cumulants of $X$ and $Y$ models. 
 The following three asymptotic regimes known from the spectral theory of random matrices 
 are also distinguished in present context. We refer to them as to 
 
 \begin{itemize}
 \item the  "full random graphs" regime, when $p= Const$ as $n\to\infty$;
 
\item  the "dilute" regime, when $p= c_n/n$ and $1\ll c_n\ll n$;
 
\item  the "sparse" regime, when $p=c/n$ and $c= const$ as $n\to\infty$.
\end{itemize}
 
\noindent 
One can consider also the fourth asymptotic regime of 
 the "very sparse" random graphs, when $p=c_n/n$ and $c_n\to 0$ as $n\to\infty$. In this regime, the diagrams of the 
leading contribution to cumulants  
are degenerated, so we formulate corresponding results as remarks to the proofs of the main theorems.
We omit subscripts in $c_n$ when they are not necessary. 

\subsection{Asymptotic behavior of cumulants}

In this subsection we formulate the main results on the asymptotic behavior of the cumulants 
$Cum_k(U_n)$. We start with the  $Y$-model.
In what follows, we denote by $u_n \asymp v_n$ asymptotic relation $u_n=v_n(1+o(1))$ as $n\to\infty$.
 
 \v
 \v
 {\bf Theorem 4.1} 
 {\it 
 Asymptotic behavior of $Cum_k(Y_n^{(q)})$ is determined by the following classes of diagrams:
 
 A) in the full random graphs regime, the leading contribution to the sum (3.5) is given by the diagrams
 $\cd_k^{(1)}(Y)$
 that have exactly  $k-1$ arcs;  the graph $G(\d_k), \d_k\in \cd_k^{(1)}(Y)$ is a tree; then 
  $$
 Cum_k(Y_n^{(q)}) \asymp \sum_{\d_k \in \cd_k^{(1)}(Y)} W(\d_k)\ n^{(q-1)k+2} \,
  \eqno (4.1)
 $$
 where $W(\d_k)$ is given by (3.6) considered with $ \d_k \in \cd_k^{(1)}(Y^{(q)})$;
 \v  
 B) in the dilute random graphs regime, the leading contribution to (3.5)  is given by 
the same diagrams as in (A), $ \d_k\in \cd_k^{(2)}(Y^{(q)}) =  \cd_k^{(1)}(Y^{(q)})$, and  
$$
Cum_k(Y^{(q)}_n) \asymp  \sum_{\d_k \in \cd_k^{(1)}(Y)} (\E a)^{(q-1)k+1}\ n^{(q-1)k+2} \, ;
\eqno (4.2)
$$

C) in the sparse random graphs regime,  the leading contribution to (3.5) is given by the
 diagrams $ \cd^{(3)}(Y^{(q)})$ such that the graph $G(\tilde \d_k), \tilde \d_k \in \cd^{(3)}(Y^{(q)})$ 
is a tree $T_l$ with $l$ edges, $1\le l \le (q-1)k+1$; then 
$$
 Cum_k(Y^{(q)}_n ) \asymp \sum_{l=1}^{(q-1)k+1} \sum_{\d_k \in \cd ^{(3)}(Y), G(\d_k) = T_l}
  (\E a)^l n^{l+1}.
  \eqno (4.3)
$$

 }

\v {\it Corollary of Theorem 4.1.} Remembering that $\E a = p = p_n$, we can reformulate the results of Theorem 4.1
in the following form:
$$
\lim_{n\to\infty}\  
{1\over p_nn^2}\, Cum_k\left({1\over (np_n)^{q-1}}Y^{(q)}_n\right) = F^{(q)}_k(\omega), \quad \omega = 1,2,3,
\eqno (4.4)
$$
where the numbers $F^{(q)}_k(i)$ represent the contributions  of corresponding families of diagrams 
(weighted by  $W(\d_k)$ in the full random graphs regime). We discuss relation (4.4) in more details in subsection 4.3.  Explicit expressions for some of $F_k^{(q)}(\omega)$ will be obtained in Section 6.

\vskip 0.5cm 
Let us consider the $X$-model. 
Now the difference between the cases of the even and odd numbers $q$ becomes crucial.

 \v
 {\bf Theorem 4.2} 
 {\it 
 Asymptotic behavior of $Cum_k(X_n)$ is determined by the following classes of diagrams:
 
 A) in the full random graphs regime, the leading contribution to the sum (3.5) is given by the diagrams
 $\cd_k^{(1)}(X^{(q)})$
 that have exactly  $k-1$ arcs; 
 in this case 
 $$
 Cum_k(X_n^{(q)}) \asymp \sum_{\d_k \in \cd_k^{(1)}(X)} W(\d_k)\ n^{(q-2)k+2}\, ,
 \eqno (4.5)
 $$
 where expressions for $W(\d_k)$ is given by (3.6) with   $ \d_k \in \cd_k^{(1)}(X_n^{(q)})$; 
\v  
 B) in the dilute random graphs regime, the leading contribution to (3.5) in the case of $X$-model 
 with even $q=2q'$is given by 
the diagrams $\cd_k^{(2)}(X)$ such that the graph 
$G(\d_k'), \d_k'\in \cd_k^{(2)}(X)$ is a tree with the maximal possible  number of edges; then 
$$
Cum_k(X^{(2q')}_n) \asymp  
 \sum_{\d_k \in \cd_k^{(2)}(X^{(2q')})} (\E a)^{(q'-1)k+1}\ n^{(q'-1)k+2};
 \eqno (4.6)
 $$
 in the case of $X$-model with odd $q$ the leading contribution to (3.5) is given by the diagrams
 $\tilde \cd^{(2)}_k(X)$ such that the graph $G(\tilde \d_k'), \tilde \d'_k \in  \tilde \cd^{(2)}_k(X)$ is a cycle with he maximal possible number of edges; then 
$$ 
Cum_k(X^{(2q'+1)}_n) \asymp  
  \sum_{\d_k \in \tilde \cd_k^{(2)}(X^{(2q'+1)})} (\E a)^{2q'+1}\ n^{2q'+1};
\eqno (4.7)
$$ 

C) in the sparse random graphs regime,  the leading contribution to (3.5) in the case of even $q=2q'+1$ is given by the
 diagrams $ \cd^{(3)}(X)$ such that the graph $G(\d_k''), \d_k'' \in \cd^{(3)}(X)$ 
is a tree $T_l$ with $l$ edges, $1\le l \le (q'-1)k+1$; then 
 $$
 Cum_k(X^{(2q')}_n ) \asymp \sum_{l=1}^{(q'-1)k+1} \sum_{\d_k \in \cd ^{(3)}(X), G(\d_k) = T_l}
  (\E a)^l n^{l+1};
  \eqno (4.8)
 $$
 in the case of $X$-model with odd $q=2q'+1$, the leading  contribution to (3.5)
 is given by the diagrams $\tilde \cd^{(3)}(X)$ such that $G(\d_k)$ is a graph $T^{(1)}_l$ with $l$ edges and one cycle only; then
 $$
 Cum_k(X^{(2q'+1)}_n ) \asymp \sum_{l=1}^{(q'-1)k+1} \sum_{\d_k \in \cd ^{(3)}(X), G(\d_k) = T^{(1)}_l}
  (\E a)^l n^{l}.
  \eqno (4.9)
 $$
 }
 
 \vskip 0.5cm 
 {\it Corollary of Theorem 4.2.}  
 In the full random graphs regime,
 $$
 \lim_{n\to\infty} \ {1\over p n^2} \ Cum_k\left( {1\over p^{q-1} n^{q-2} } X^{(q)}_n\right) = 
 \Phi^{(q)}_k(1),
 \eqno (4.10)
 $$
 and in the dilute and sparse regimes we have, respectively, 
$$
\lim_{1\ll c\ll n } \ {1\over c n}\  Cum_k \left( {1\over c^{q'-1}} X_n^{(2q')}\right) =
\Phi^{(2q')}_k(\omega), \quad \omega = 2,3,
\eqno (4.11)
$$
and 
$$
\lim_{1\ll c\ll n} {1\over c^{2q'+1}} Cum_k(X^{(2q'+1)}_n) = \Phi_k^{(2q'+1)}(2)
\eqno (4.12)
$$
and 
$$
\lim_{n\to\infty, c= const} Cum_k(X^{(2q' +1)}_n) = \Phi_k^{(2q'+1)}(3)
\eqno (4.13)
$$
We discuss relations (4.10)-(4.13) in Subsection 4.3. Explicit expressions for some of $\Phi_k^{(q)}(\omega)$ will be obtained in Section 6.

\subsection{Proof of Theorems 4.1 and 4.2}

{\it Proof of Theorem 4.1.} In the  full random graphs regime, $\E a$ is a constant and all the terms of the sum (3.6) are of the same order of magnitude. The leading contribution to (3.5) comes from the diagrams 
$\d_k$ such that the graph $G(\d_k)$ has maximally possible number of vertices. In this case the number of arcs in $\d_k$ is minimal, i.e. is equal to $k-1$ and  $G(\d_k)$ is a tree.
This proves (4.1). 

In the case of dilute and sparse graphs, relation $\E a^2 = o(\E a)$ shows that the leading contribution to the sum (3.6) is obtained from those partitions $\pi_s$ that  $\chi(\pi_s;\d_k) =0$. 
Since $\d_k$ is connected, then  only trivial partition $\pi_0$ verifies this condition. 
Let us denote by $l$ the number of edges in $G(\d_k)$. If $G(\d_k)$ is given by a tree, then $\nu(\d_k) = l+1$ and 
$(\E a)^l \vert \cn (\d_k)\vert  = n c^l(1+o(1))$. If $G(\d_k)$ is not a tree, then by the Euler theorem, $\nu(\d_k)< l$ and such 
diagrams provide a contribution to (3.5) of the order $o(nc^l)$. In the case of dilute random graphs,
$c\to\infty$ and the leading contribution is obtained from the graphs with$l= (q-1)k+1$; other trees do not contribute. This proves (4.2). To show (4.3), it remains to note that  in the sparse random graphs regime all trees
with $1\le l\le (q-1)k+1$ provide contributions of the same order of magnitude. Theorem 4.1 is proved.
\vskip 0.5cm
{\it Proof of Theorem 4.2}. In the full random graph regime, the leading contribution to (3.5) comes from 
those $\d_k$ that have $k-1$ arcs. This implies relation (4.5).

It is easy to see that in the dilute and sparse regimes, the only trivial partition $\pi_0$ 
contributes to (3.6). In the case of $X^{(2q')}$, we repeat arguments of the proof of Theorem 4.1 and get relations (4.6) and (4.8).

Let us pass to  the case of odd $q=2q'+1$
and consider  diagrams $\d_1$ with one element $\l_1$. It is clear that  
graphs $G(\d_1)$ always contain at least one cycle. Then $G(\d_k)$ also contain at least one cycle.
If the number of edges of $G(\d_k)$ is equal to $l$, then the contribution 
of such a diagram is of the order $(\E a)^l n^{l+1-w}$, where $w$ is the number of cycles in $G(\d_k)$.
The graphs with one cycle only provide the leading contribution of the order $(\E a)^l n^l$.  
In the dilute random graphs regime the leading contribution comes from the graphs
with maximal number of edges. Then we conclude that in this case $G(\d_k)$ represents 
a cycle with $2q'+1$ edges and $2q'+1$ vertices. This proves (4.7). In the case of sparse regime,
we consider $\d_k$ such that $G(\d_k)$ has exactly one cycle. This implies (4.9).
Theorem 4.2 is proved.

\vskip 0.5 cm
{\it Remark.} It follows from the proof of theorems 4.1 and 4.2 that 
if $p_n = c/n$ and $c\to 0$ as $n\to\infty$, then 
the leading contribution to $Cum_k(Y^{(q)}_n)$ and $Cum_k(Y^{(q)}_n)$ is given by the connected diagrams $\d_k$ such that $G(\d_k)$ have minimal number of edges. 
It is easy to see that in this case 
$$
Cum_k(Y^{(q)}_n) = 2^{k-1} p_n n^2 (1+o(1)) = 2^{k-1}cn(1+o(1)),
\eqno (4.14)
$$
and
$$
Cum_k(X^{(2q')}_n) = 2^{k-1} p_n n^2 (1+o(1)) = 2^{k-1}cn(1+o(1)).
\eqno (4.15)
$$
Also we have $Cum_k(X^{(2q'+1)}_n) = O(p_n^3 n^3) = O(c^3)$ as $n\to\infty, c\to 0$.

\subsection{Central Limit Theorem for variables $X$ and $Y$} 

Regarding the definition of the cumulants, 
it is easy to see that for any constant $C$,
$$
Cum_k(V_n+C) = Cum_k(V_n) + C\d_{k,1}.
$$
Also, if there exists such a sequence $b_n $  that 
$$
{1\over b_n} Cum_k(V_n) \to \phi_k, \quad k\ge 1,
\eqno (4.16)
$$
then 
$$
Cum_k \left( {V_n - \E V_n\over \sqrt {b_n}} \right) \to \cases{ \phi_k, & if $k=2$; \cr
0, & if $k\neq 2$.\cr }
\eqno (4.17)
$$
Relation (4.17) means that the probability law of the centered  random variable 
$\tilde V_n = {(V_n - \E V_n)/\sqrt{b_n}}$ converges  to the  Gaussian 
distribution ${\cal N}(0, \phi_2)$. 
Regarding centered and normalized variables $X^{(q)}_n$ and $Y^{(q)}_n$, we formulate the Central Limit Theorem. 

\v
{\bf Theorem 4.3}  
{\it   If $n\to\infty$, and $p_n$ determines one of the three asymptotic regimes indicated by $\omega$, 
the following random variables converge in law to the Gaussian distribution:
$$
{1\over \sqrt { p_n n^2} }\cdot  {Y_n^{(q)} - \E Y_n^{(q)}\over  (p_n n)^{q-1}} \to  {\cal N}(0, F^{(q)}_2(\omega)), \quad \omega=1,2,3,
\eqno (4.18)
$$
and  
$$
{1\over \sqrt { p_n n^2} }\cdot  { X_n^{(2q')} - \E X_n^{(2q')}\over (p_n n)^{q'-1}} \to  {\cal N}(0, \Phi^{(2q')}_2(\omega)), \quad \omega=2,3.
\eqno (4.19)
$$
Also convergence to the standard normal distribution holds in the case of $X$ model in the full random graphs asymptotic regime;
$$
{1\over \sqrt { p_n n^2} } \cdot { X_n^{(q)} - \E X_n^{(q)} \over (p_n n)^{q-2}}\to  {\cal N}(0, \Phi^{(q)}_2), 
\eqno (4.20)
$$
and in the regime of dilute random graphs for $X^{(q)}$ with odd $q= 2q'+1$:
$$
{1\over c^{q'+1/2}} \left (X_n^{(2q'+1)} - \E X_n^{(2q'+1)} \right) \to  {\cal N}(0, \Phi^{(2q'+1)}_2(B)).
\eqno (4.21)
$$
 There is no convergence to the normal distribution  of $X_n^{(2q'+1)}$ 
in the regime of sparse random graphs. }

\v
{\it Proof.} The proof follows immediately from relations (4.4) and (4.11) with properties (4.16) and (4.17) taken into account.

\v
Let us discuss relations of Theorem 4.3 with the spectral theory of random matrices. 
Remembering that $p_n n= c$ and introducing the normalized adjacency matrices 
$$
\hat A_{(n,c)} = {1\over \sqrt c} A_n,
$$
we can introduce variables $M_{2q'}^{(n,c)}= {1\over n} \T \hat A_{(n,c)}^{2q'}$
and 
rewrite (4.19) in the form 
$$
\sqrt{ {n c}} \left( M_{2q'}^{(n,c)} - \E M_{2q'}^{(n,c)}\right) \to {\cal N}(0, \Phi^{(2q')}_2(\omega)), \quad \omega=2,3.
\eqno (4.22)
$$
Also, we deduce from relations (4.11)-(4.13)  that 
$$
\lim_{n\to\infty}  \E M_{2q'}^{(n,c)} = \Phi_1^{(2q')}(\omega) = m_{2q'} (\omega)
\ \  {\hbox {
and}}\ \  \lim_{n\to\infty}  \E M_{2q'+1}^{(n,c)}=0.
\eqno (4.23)
$$

Variables $M_{q}^{(n,c)}$ represent the moments of the normalized eigenvalue counting measure
of random matrices $\hat A_{(n,c)}$,
$$
\sigma_n(\lambda) = \# \{ j: \lambda_j^{(n,c)}\le \lambda\} n^{-1}.
$$
Convergence (4.23) implies the weak convergence of the measures $\sigma_n$. This convergence is 
established in  the present and more general settings in 
papers \cite{BG,KV} and \cite{KSV}. Central Limit Theorem (4.22) improves results of these papers.

\v 
Let us note that relations (4.14) and (4.15) imply that 
$$
{1\over cn}Cum_k(Y^{(q)}_n) \to 2^{k-1} \quad {\hbox{as}}\ \ n\to\infty, c\to 0.
\eqno (4.24)
$$
The same is true for the cumulants of $X^{(2q')}_n$. This means that 
the Central Limit Theorem does not hold for these variables  in the asymptotic regime of very sparse random graphs. 

Returning to variables $Y_n^{(q)}$, we rewrite (4.4) with $k=1$ in the form
$$
\lim_{n\to\infty } {1\over n c^q} \ \E \, Y_n^{(q)} = F_1^{(q)}(\omega)=1,
\eqno (4.25)
$$ 
since there exists only one connected diagram. Variable $Y_n^{(q)}$ counts the number of $q$-step walks over the graph, and relation (4.25) shows that this number is asymptotically proportional to $nc^q$. It  is natural to expect this result  because the average degree of a vertex in the  Erd\H os-R\'enyi random graph
converges to $c$, as $n\to\infty$ \cite{JLR}.
It follows from (4.18) that 
$$
{1\over nc^q} Y_n^{(q)} - \E \left\{{1\over nc^q} Y_n^{(q)}\right\} \sim {\gamma_n\over \sqrt{nc}},
$$
where $\gamma_n$ converges in law to a gaussian random variable. 
Then convergence with probability 1 holds
$$
{1\over nc^q} Y_n^{(q)}  \to 1
$$
 as $n\to\infty$ in the dilute and sparse random graphs regimes.
The limiting variance of $\gamma_n$ depends on the asymptotic regime. We return to this question in Section 6.

\section{Number of tree diagrams}

In this section we derive recurrent relations that determine the number 
of connected diagrams $ \d_k = (\L_k,\Sigma)$ with minimal number of arcs,
$$
\cd^{(1)}_k = \{  (\L_k,\Sigma): \  \vert \Sigma\vert = k-1\}.
$$
In the case of $Y$-model, the graphs $G(\d_k)$ have the tree structure and we refer to $\d_k$ as to the tree diagrams.  In the case of $X$ model we refer to  such diagrams as to the tree-like diagrams. It is clear that $$\vert \cd^{(1)}_k(X^{(q)})\vert = \vert \cd^{(1)}_k(Y^{(q)})\vert = d^{(q)}_k,$$ so we consider the case of the $Y$-model only.

\subsection{Recurrent relations and Lagrange equation}

 First let us remind that the arcs in $\d_k$ are interpreted by $G(\d_k)$ as gluing
between edges that make a color group. 
Let us say that the edges that are not glued 
are not colored and stay grey. So, we forget the color of the simple color groups consisting of one edge only.
The different color groups and grey edges correspond to the edges of $G(\d_k)$. 
It is convenient to color   the arcs that join the edges of one group in the same color as the edges of this group.
We start with the following simple statement.

\vskip 0.5cm 
{\bf Lemma 5.1.} {\it Let a  diagram $\d_k\in \cd^{(1)}_k(Y^{(q)})$ have arcs of $s$ different
colors. 
Then  there are $(q-1)k - s+1$ grey edges in $\d_k$.}

\vskip 0.3cm
{\it Proof.}  Let us assume that   there are $\mu_j$ arcs of each color.
 Obviously, $\mu_1+\dots+\mu_s = k-1$ and the total number
of colored edges is $k-1+s$. Taking into account that the total number of edges in $\d_k$ is $qk$, we obtain the result. Lemma is proved. 

\vskip 0.5cm 
{\bf Lemma 5.2} {\it Given $q\ge 2$, the numbers $d_k^{(q)}= \vert \cd^{(1)}_k(Y^{(q)})\vert $ with $ k\ge 1$
are given  by equalities 
$$
d^{(q)}_k = { k!\over(q-1)k+1} h^{(q)}_k,
\eqno (5.1)
$$
where the sequence $\{ h_k^{(q)}\}_{k\ge 0}$ is 
determined by the recurrent relation 
$$
h_k^{(q)} = {(q-1)k +1\over k}\cdot \left( 
\sum_{{\stackrel{j_1+\dots + j_q=k-1} {j_i\ge 0}} }  h_{j_1}^{(q)} \cdots h_{j_q}^{(q)}\right), \  k\ge 1 
\eqno (5.2)
$$
with initial condition $h^{(q)}_0=1$.} 

\vskip 0.5cm 

{\it Proof.}  Let us remind that since our diagrams are reduced, then each color group
has a unique maximal edge.
Now let us count the number of the diagrams $\bar \d_k$ such that the last element $\l_k$ contains one and only one  edge that is the   
maximal edge of  a color group. This means that only one arc ends at $\l_k$ by its right foot (or leg). 
Let us denote this arc by $\bar \s$.
\vskip 4cm

\begin{figure}[htbp]
\centerline{\includegraphics[width=12cm]{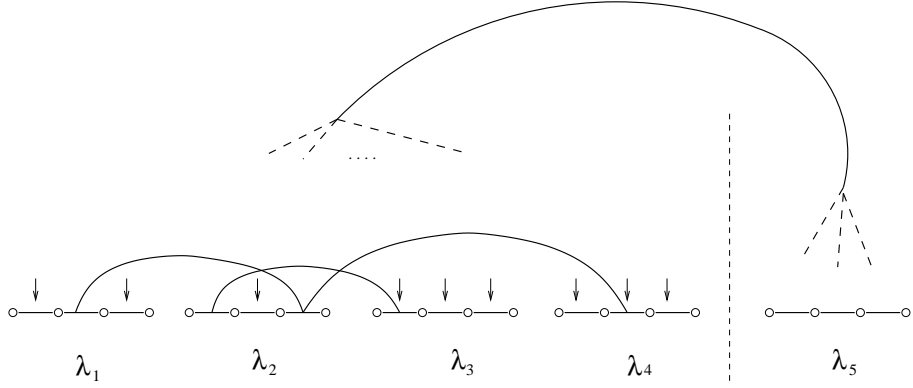}}
\caption{\footnotesize{Example of $\bar \d_k$ with possible positions for for the left foot of  $\bar \s$} }
\end{figure}

Obviously, 
one can choose one of the $q$ edges of $\l_k$ to put this right leg. The left leg of $\bar \s$ joins $\l_k$ with connected tree  diagram  $\d_{k-1} \in \cd^{(1)}_{k-1}(Y^{(q)})$. By Lemma 5.1, there are $(q-1)(k-1) -s +1$ grey edges and $s$ maximal edges of $s$ color groups, where one can put the left foot of $\bar \s$. 
Then we obtain relation
$$
\vert \bar \cd_k^{(1)}\vert = \left((q-1)(k-1)^{ } +1 \right) d_{k-1}^{(q)}.
$$

Now let us consider the case when there are $l$ arcs $\bar \s_1, \dots , \bar \s_l$ 
that  end at $\l_k$ with \mbox{ $2\le l\le \min(q,k-1)$.} There is ${q\choose l}$  possibilities to choose the emplacements for the right legs  of these arcs.  To put the left legs of these  arcs, one has to choose the $l$ subsets $\L^{(1)}, \dots,\L^{(l)}$  of $j_i = \vert \L^{(i)}\vert$ elements such that 
 $j_1+\dots+ j_l = k-1$ and  $ j_i\ge 1$, then  to create $l$ connected sub-diagrams $(\L^{(i)}, \Sigma^{(i)}), i=1,\dots, l$ and to choose the emplacements 
for the left legs of $\bar \s_i$ in $(\L^{(i)}, \S^{(i)})$. This produces 
$$
{q \choose l} \sum_{\stackrel{j_1+\dots +j_l=k-1} {j_i\ge 1} } {(k-1)!\over j_1!\cdots j_l!}
\ \prod_{i=1}^l ((q-1)j_i +1) d^{(q)}_{j_i}
$$
diagrams. Summing up, we derive for numbers $\{d^{(q)}_k,  k\ge 1\}$ with given $q\ge 2$ the following  recurrent relation 
$$
d_k^{(q)} = \sum_{l=1}^q I_{  [1, k-1] }(l)\times   {q \choose l} 
\sum_{\stackrel{j_1+\dots +j_l=k-1}{ j_i\ge 1}}{(k-1)!\over j_1!\cdots j_l!}\  \prod_{i=1}^l ((q-1)j_i +1) d^{(q)}_{j_i},
\eqno (5.3)
$$
with initial condition $d^{(q)}_1=1$. Here we denoted by $I_{[1,k-1]}(\cdot)$ the indicator function of the interval $[1,k-1]$.

Introducing the auxiliary numbers $h_j$, such that 
$$
h_j = {(q-1)j+1 \over j!}d^{(q)}_j, \quad j\ge 1, 
$$
 we reduce (5.3) to relation
$$
{k\over (q-1)k+1} h_k = \sum_{l=1}^q  I_{  [1, k-1] }(l)\times  {q \choose l} 
\sum_{\stackrel{j_1+\dots +j_l=k-1} {j_i\ge 1}} h_{j_1} h_{j_2} \cdots h_{j_l}, \ k\ge 1.
\eqno (5.4)
$$

If we introduce an auxiliary number $h_0=1$, then we can rewrite (5.4) in more compact form. Indeed, 
we can make an agreement  that each labeled  edge $\varepsilon ^{(k)}_l$ of the element $\l_k$ 
serves as the right foot of 
the corresponding arc
$\bar \s_l$ but some of these arcs can have the right leg "empty", i.e. with no elements attached to their right feet. Then the variable $l$ of  the sum of (5.4) represents the number of non-empty arcs, and 
${q \choose l}$ stands for the choice of these non-empty arcs from the set $\bar \s_1, \dots, \bar\s_q$. 
This corresponds to the choice of the variables $j_i$ in the product $h_{j_1}\cdots h_{j_q}$ that take zero value.
Finally, we set $h^{(q)}_j = h_j, j\ge 0$ and  conclude that relation (5.4) is equivalent to recurrent relation (5.3). Lemma 3.5 is proved.

\vskip 0.5cm
{\bf Lemma 5.3.}

{\it The generating function 
$$
H_q(z) = \sum_{k=0}^{+\infty} h_k^{(q)} z^k
\eqno (5.5)
$$
is regular in the domain $\{ z: \vert z\vert \le 1/(q^2e)\}$ and 
verifies there the analog of the equation 
$$
H_q(z) = \exp\{ qz \left(H_q(z)\right)^{q-1}\}.
\eqno (5.6)
$$
It follows from (5.3) that the numbers $h^{(q)}_k$ can be  found from 
recurrent relations
$$
\sum_{\stackrel{j_1+ \dots + j_{q-1} = k}{ \ j_i \ge 0}} h^{(q)}_{j_1} \cdots h^{(q)}_{j_{q-1}} = q^k(q-1)^k {(k+1)^{k-1}\over k!}, \quad h^{(q)}_0=1.
\eqno (5.7)
$$
}
{\it Remark.} Relation (5.6) with $q=2$ is known as the Lagrange (or P\'olya) equation \cite{L,PS,S}. In the next subsection we  give another derivation of (5.6) by using
 the notion of color trees.
 In Section 6 we obtain  one more generalization of the Lagrange equation. 

\vskip 0.3cm 
{\it Proof of Lemma 5.3.}
It is easy to deduce from (5.2) that $H_q(z)$ verifies 
integral equation 
$$
H_q(z)-1 = z(q-1) (H_q(z))^q + \int_0^z (H_q(\zeta))^q \ d\zeta .
$$
that is  equivalent to the differential equation 
 $$
 {dH_q(z)\over dz} = {q\left(H_q(z)\right)^q\over 
 1 - (q^2-q)z \left(H_q(z) \right)^{q-1}}, \quad H_q(0)=1.
 \eqno (5.8)
 $$
Substitution 
$$
\psi(z) = z \left( H_q(z)\right)^{q-1}
\eqno (5.9)
$$ 
transforms (5.8)  to an elementary equation
$$
\psi'(z) = {1\over z} \cdot { \psi(z)\over 1 - (q^2-q) \psi(z)}.
$$
Resolving this equation with obvious initial condition,
we conclude that $\psi(z)$ verifies the algebraic equation 
$$
 \psi(z) = z e^{q(q-1)\psi(z)}
 \eqno (5.10)
 $$
 known as the P\'olya equation. 
Then (5.6) easily follows from (5.9) and (5.10).

Using the standard method of the contour integration, we deduce from (5.10) explicit 
expressions  for the coefficients $\psi_k$ 
 of the generating function $\psi(z) = \sum_{k\ge 0} \psi_k z^k$. 
The first observation is that  the inverse function $\psi^*(w) = we^{-q(q-1)w}$ is regular in the vicinity of the origin.
By the Cauchy formula, we have
$$
\psi_k = {1\over 2\pi \i} \oint {\psi(z)\over z^{k+1}} dz.
$$
By changing variables by $z = \psi^*(w)$, we get $ dz = (1-q(q-1)w)e^{-q(q-1)w} dw$ and
find that
$$
\psi_k = {1\over 2\pi \i} \oint {1-q(q-1)w\over w^k} e^{q(q-1)wk} dw.
$$
Then 
$$
\psi_k = (q(q-1))^{k-1} {k^{k-2}\over (k-1)!},
\eqno (5.11)
$$
and  (5.7) follows. Lemma 5.3 is proved.

\subsection{Color trees}

In this subsection we give another derivation of the Lagrange equation (5.6) based on the graph representation $G(\d_k)$ of the diagrams $\d_k$. For simplicity, we consider the case of $Y$-model with $q=2$ only.

If $\d_k$ has $k-1$ arcs, then $G(\d_k)$ is a tree. We say that the arcs that indicate the groups of edges to be glued
are all of  the same color. Then the corresponding edges of $G(\d_k)$ are of this color. 
We say that 
the edges that are not glued are grey. 

Let us derive expressions for the number  of the  elements of the set  
$$
T_k = \vert {\cal G}_k\vert, \ \  {\cal G}_k = \{ G(\d_k), \d_k \in \cd^{(1)}(Y^{(2)})\}.
$$ 
We denote by $\h T_m$  the number of rooted trees of the form $G(\d_m)$.

Let us consider left element of  $\l_1$ as the root edge $\rho$
for the trees we construct (see Figure 4). 
We can attach to this root $r$ elements $\nu_1,\dots \nu_r$ that we choose by ${m\choose r}$ ways. 
We glue these elements by their right edges. 

\begin{figure}[htbp]
\centerline{\includegraphics[width=12cm]{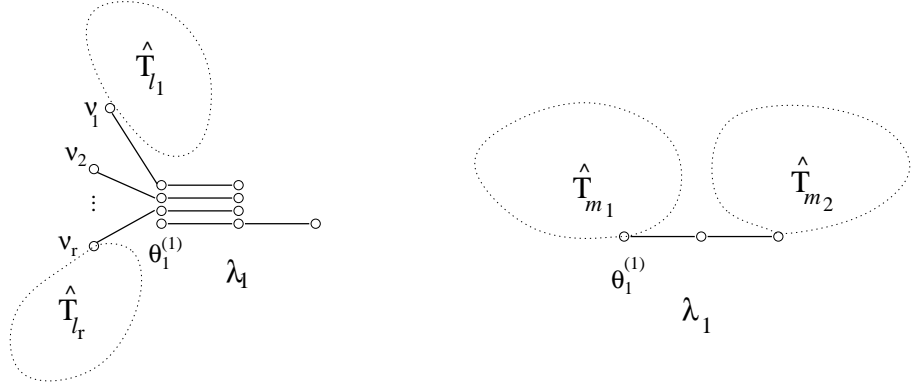}}
\caption{\footnotesize{Rooted color tree and color tree} }
\end{figure}

Then we separate $m-r$ elements into the groups of $l_1$, $l_2, \dots, l_r$ elements and 
construct rooted trees and attach  them by their roots to the free elements of $\nu_1,\dots ,\nu_r$.
There are ${(m-r)!\over l_1!\cdots l_r! }$  possibilities to separate $m-r$ elements into these groups.
Then we get 
$$
\h T_m = \sum_{r=1}^m 2^r {m\choose r} \sum_{l_1+\dots +l_r= m-r} {(m-r)!\over l_1!\cdots l_r!} \h T_{l_1} \cdots \h T_{l_r}, \quad m\ge 1,
\eqno (5.12)
$$
where the sum runs over all $l_i\ge 0$ and we  accepted that  $\h T_0=1$.

Simplifying (5.12) and taking into account that $\h T_0 = 1$, we derive from (5.12) the following relation for the generating function $t(z) = \sum_{k\ge 0} t_k z^k$, $t_k = \h T_k /k!$ (cf (5.6)):
$$
t(z) = \exp\{ 2z t(z)\}.
\eqno (5.13)
$$
Then $\h T_m = 2^m (m+1)^{m-1}$ and the numbers $t_m$ are given by  recurrent relations
$$
t_m = {m+1\over m} \sum_{j=0}^{m-1} t_j t_{m-1-j}, \quad t_0=1.
\eqno (5.14)
$$

Now let us return to the construction of the tree $G(\d_k)$. Now the left edge of $\l_1$
can serve as the root for the left rooted tree of $m_1$ elements and the right element
serve as the root for the rooted tree of $m_2 = k-1-m_1$ elements. Then
$$
T_k = \sum_{\stackrel{m_1+ m_2 = k-1}{m_i\ge 0}} {(k-1)!\over m_1! m_2!}\  \h T_{m_1} \, \h T_{m_2},
$$
where  $\h T_0 = 1$. Then
$$
{T_k\over (k-1)!} =  \sum_{j=0}^{k-1} t_j t_{k-1-j} = {k\over k+1} t_k = {2^k (k+1)^{k-2}\over (k-1)!}.
$$
Then equality $T_k= 2^k(k+1)^{k-2}$ follows.

\vskip 0.5cm 
In the general case $q\ge 2$, one can easily repeat this procedure of tree construction and derive (5.6) directly, 
without use of the numbers $h^{(q)}_k$.

\section{Limits of the cumulants}

In this section we specify the limits  of the cumulants of the $Y$-model
$F_k^{(q)}(i)$ (4.4) and of the $X$-model $\Phi_k^{(q)}(i)$ (4.10)-(4.13). 
We start with the case of full random graphs regime, but the most attention is paid to the dilute and sparse random graphs regimes. 

\subsection{Cumulants of full $X$ and $Y$ models}

In Section 4 we have shown that the leading contribution to the cumulants of (4.1)
and (4.5) is given by those diagrams $\d_k$ that have exactly $k-1$ arcs and the graph $G(\d_k)$ 
with the maximal number of vertices. The weight of the diagram $W(\d_k)$ is determined by relation (3.6),
where all terms  $(E a)^l$
are of the same order of magnitude (see the proof of Lemma 3.2). 
 Regarding $Y$-model and taking into account that in this case 
$m(\d_k) = (q-1)k +1 $ (see Lemma 5.1), we combine (3.6) with (4.1) and 
conclude  that (4.4) is true with
$$
F^{(q)}_k(1) = \sum_{\d_k\in \cd^{(1)}_k(Y)} \ \sum_{\pi_s\in \Pi_k} (-1)^{s-1} (s-1)! \, p^{\chi(\pi_s;\d_k)},
\eqno (6.1)
$$
where $\chi(\pi_s;\d_k)$ is determined in Lemma 3.2. 

Considering $X$-model, it is easy to see that (3.6) gives the same expression for $W(\d_k)$ as for the $Y$-model.  
Using (4.5), we arrive at the conclusion that (4.10) is true with 
$\Phi^{(q)}_k(A) = F^{(q)}_k(A)$ (6.1).

Restricting ourself to  the first two cumulants, it is easy to compute that
$$
F^{(q)}_1(1) = \Phi^{(q)}_1(1) = 1
$$ 
and that
$$
F^{(q)}_2(1) = \Phi_2^{(q)}(1) =  2q^2(1-p).
$$

\subsection{Cumulants of dilute  $Y$-model}

Regarding (4.2) with $\E a = c/n$ and using results of Section 5, 
we conclude that (4.4) takes the form of
$$
F_k^{(q)}(2) = \lim_{n,c\to\infty} {1\over nc} Cum_k \left( {1\over c^{q-1} }Y_n^{(q)}\right) = 2^{k-1}\vert \cd_k^{(1)}(Y)\vert = 2^{k-1} d^{(q)}_k,
\eqno (6.2)
$$
where $d_k^{(q)}$ are determined by relations (5.1) and (5.2). In (6.2), we have taken into account 
that the $k-1$ arc of the diagrams $\d_k$ can be drawn in the direct and inverse sense. This produces the factor $2^{k-1}$.

Let us consider in more details the case of $q=2$ that corresponds to the 
continuous matrix model  (2.1) with the quartic potential.   
In this case relation (5.2) takes the  form
$$
h^{(2)}_k = {k+1\over k} \sum_{j_1+j_2=k_1, j_i\ge 0} h^{(2)}_{j_1} h^{(2)}_{j_2}, \quad h^{(2)}_0=1.
$$
It follows from (5.7) that in this case
$h^{(2)}_k = {2^k  (k+1)^{k-1}/ k!} $
and
$$
  d^{(2)}_k= 2^k (k+1)^{k-2}.
\eqno (6.3)
$$
The combinatorial meaning of $d^{(2)}_k$ can be  explained with the help of the notion of the dual diagrams that we briefly describe below.

\begin{figure}
\centerline{\includegraphics[width=12cm]{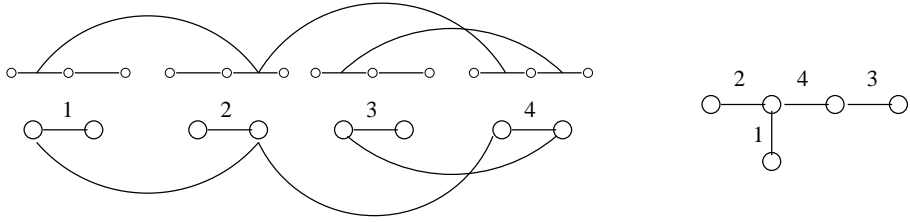}}
\caption{\footnotesize{ Dual diagrams and dual trees in the case of $q=2$, $k=4$}}
\end{figure}

 Regarding $\d_k \in \cd^{(1)}(Y^{(2)})$, we construct the dual diagram $\hat d_k$ by  transforming  the edges of elements $\l_j$ into vertices joined by an edge; the arcs remain without changes and indicate the vertices of $\hat \d_k$ to be glued (see \mbox{Figure 5}). 
It is easy to see that  the graph $G(\hat \d_k)$ is a tree. Then $d^{(2)}_k$ is equal to the number of 
all non-rooted trees constructed with the help of $k$ labeled edges. Since the edges of $\d_k$ are labeled to be the left and the right one,  we can think about the edges of $\hat G(\d_k)$ are the oriented
one. 

Regarding the dual diagrams in the general case of $q\ge 2$, we see that 
these are trees when $q=2$ and trees constructed from oriented chains of $q$ edges when $q>2$. It is not clear, whether it is possible to derive from (5.7)  explicit expressions for $d^{(q)}_k$ with  $q>2$. 

\subsection{Cumulants of sparse $Y$-model}

The diagrams of $\cd^{(3)}(Y)$ are of more complicated structure than those of $\cd^{(1)}(Y)$ and we can not obtain recurrent relations to determine $F^{(q)}_k(3)$ in the general case of $q\ge 2$ and 
$k\ge 1$. We present the results for $F^{(2)}_k(3)$ and $F^{(q)}_1(3)$ only.

\v
{\bf Lemma 6.1}  
{\it The coefficients $F^{(2)}_k(3), k\ge 1$ are determined by relations
$$
F^{(2)}_k(3) = {2^{k-1}(k-1)!\over c^k}  w_k, 
\eqno (6.4)
$$
where $w_k$ are determined by relation
$$
w_k  =  \sum_{s=1}^{k} {1\over (s-1)!} 
\, \sum_{j=0}^ {k-s} \h w_{j}\, \h w_{k-s-j}, \quad k\ge 1,
\eqno (6.5)
$$
and  the numbers $\h w_{j}, j\ge 0$ are such that the generating function $\h W(z) = \sum_{k\ge0} \h w_k z^k$ verifies equation}
$$
\h W(z) = 1 - ce^{2z} + c \exp\left\{ 2z [e^{2z} \h W(z) - 1]\right\}.
\eqno (6.6)
$$
\v
{\it Proof.} 
Following the lines of subsection 5.2, we consider first the set of the rooted trees of the type we are interested in and denote their contribution by  $\h T_m$, where $m$ is the number of 
elements $\l$. 

\begin{figure}[htbp]
\centerline{\includegraphics[width=12cm]{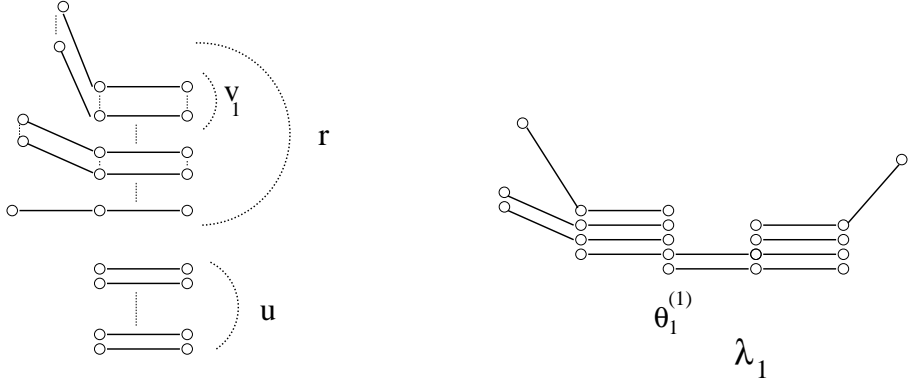}}
\caption{\footnotesize{Elements to construct a rooted tree and example of the tree}}
\end{figure}

 In the present case the trees are constructed with the help of  elements $\l$, but there can be double elements that are glued to one of the edge, and there can be simple elements,
 and there are elements glued to simple elements all the long  (see Figure 6). Let us consider the root edge $\rho$. 
 Denoting by $u$ the number of double elements attached  to $\rho$ and by $r$ the number of simple elements attached to $\rho $ by one edge. Also we denote by  
 $v_i$, $i=1,\dots, r$ the numbers of  elements glued all the long to  these $r$ simple elements.
 Taking into account possibilities to choose $u,r$, and $v_i$ elements from the set of $m$ labeled elements, we obtain the recurrent formula
$$
\h T_m = 
 c \sum_{ u=0} ^m 2^u {m \choose u} \cdot 
 \sum_{r=1}^{m-u} 2^r {m-u \choose r}
\sum_{v_1 +\dots+v_r = V,\,  V=1}^{m-u-r} 2^{v_1}\cdots 2^{v_r} {m-u-r\choose v_1,\dots,v_r} 
 $$
$$
\times \sum_{l_1+\dots +l_r = m-u-r-V} {m-u-r-V\choose l_1, \dots, l_r}\  \h T_{l_1} \cdots \h T_{l_r},
\eqno (6.7)
$$
where we denoted the multinomial coefficients
$$
{m-u-r\choose v_1,\dots,v_r} = {(m-u-r)!\over v_1!\cdots v_r!\, (m-u-r-V)!}.
$$
Factor $c$ in  the right-hand side of (6.7) takes into account the contribution (or weight) of the root $\rho$. Simplifying (6.7) and denoting $\h w_m = \h T_m/m!$ for  $m\ge 1$, we get relation
$$
\h w_m = c \sum_{ u=0} ^m {2^u\over u!} \cdot \sum_{r=1}^{m-u} {2^r \over r!}
\cdot \sum_{{V=v_1+\dots +v_r, V= 1}}^{m-u-r}  
{2^{v_1}\cdots 2^{v_r}\over v_1!\cdots v_r!}
\sum_{l_1+\dots +l_r=m-u-r-V} \,  \h w_{l_1}\cdots \h w_{l_r}.
$$
Passing to the generating function  $\h W(z)$, we derive equality
$$
\h W(z) -1 = c e^{2z}\left[ \exp\left\{ 2z e^{2z} \h W(z)\right\} - 1 \right],
\eqno (6.8)
$$
 that is equivalent to (6.6). Let us note here that (6.8) generalizes the P\'olya equation (5.13).  
 
 Returning to the set $\cd^{(3)}_k(Y)$, we consider $\l_1$ as the simple root element. Assuming that there are $s-1$ 
 elements glued all the long this root element, we dispose of  $k-s$ 
 elements to construct the rooted trees using the left and the right edge of $\l_1$ as the roots $\rho_1$ and $\rho_2$. 
 Denoting  the number $T_k = \vert \cd^{(3)}_k(Y)\vert$, we get  relation
 $$
 T_k= \sum_{s=1}^k { k-1\choose s-1}\  \sum_{\stackrel{m_1+m_2=k-s}{ m_i\ge 0}}
 {k-s\choose m_1,m_2} \hat T_{m_1}\ \hat T_{m_2}.
 $$
 Simplifying this relation and passing to variables $\hat w_m = \hat T_m/(m-1)!$, we obtain 
 equality (6.5). This completes the proof of Lemma 6.1.

\vskip 0.5cm 
In the general case of $q\ge2$ we determine the first cumulant only, 
$$
F^{(q)}_1(3) = \lim_{  \stackrel{n\to\infty, p=c/n}{c= Const} }\  \sum_{i_1,\dots, i_{q+1}=1}^n \E \left\{ A_{i_1i_2} A_{i_2i_3} \cdots 
A_{i_qi_{q+1}}\right\}.
\eqno (6.9)
$$
The vertices $\theta$  of $\l$ are ordered and we denote $ \theta_1=\rho$ and  $ \theta _2=\nu$. 
It follows from Theorem 4.1 that $F_1^{(q)}(3)$ is determined by the  number of 
diagrams $\d_1$ consisting of one element $\l$ such that corresponding graph 
$G(\d_k)$ is a tree of $l$ vertices, $2\le l\le q$. This problem is closely related with the studies done in \cite{BG,KV} and formalized in \cite{KSV}. These studies are related with the $X$ model and we present the results  in the next subsection.  In what follows, we describe briefly corresponding approach and modify it to suit  our situation. 

To study the right-hand side of (6.8), it is useful to introduce the notion of a walk $\xi$. 
A walk $\xi_q$ of $q$ steps is an ordered sequence of $q$ letters, starting  with $\rho$ and followed by $\nu$ always. 
Given a sequence $I_{q+1} = (i_1,i_2,\dots, i_{q+1})$, we construct corresponding 
$\xi_q = \xi(I_{q+1})$ by the following recurrent rule: regarding the value $i_{s+1}, s\ge 1$, we compare it with the 
values of $\{i_1, \dots, i_s\}$ and write a letter number $s+1$ in $\xi$. 
If there is no $i_j, 1\le j\le s$ such that  $i_{s+1}= i_j$, the we write a new letter that is not present in $\xi_s$. 
If there is such $j$, $1\le j\le s$ such that $I_{s+1}= i_j$, then we write on the $s+1$-th place the letter number $j$ of $\xi_s$. 

We can represent the walks graphically, by drawing the root vertex $\rho$, the next vertex $\nu$ and corresponding 
edge and then passing along $\xi_q$ and creating the new vertices at the instants,
when the new letters occur in $\xi_q$. We also draw the new edges in this case. 

\begin{figure}[htbp]
\centerline{\includegraphics[width=12cm]{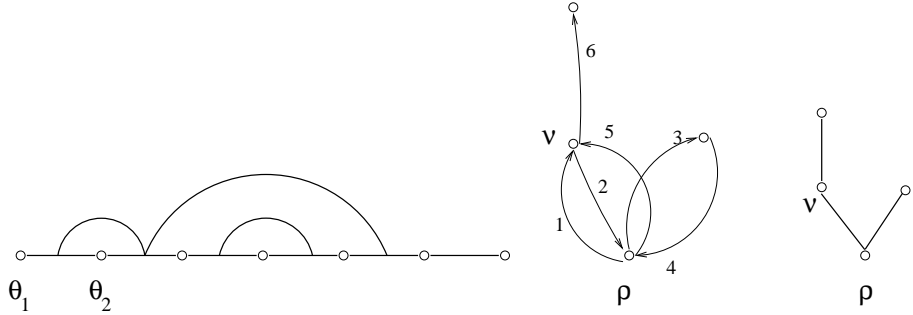}}
\caption{\footnotesize{A diagram $\d_1$, the walk $\xi_q$ and corresponding graph $G(\d_1)={\cal G}(\xi_q)$ }}
 \end{figure}

It is clear that there is one-to-one correspondence between the diagrams $\d_1$ and walks $\xi_q$. 
Also, the walk $\xi_q$ generates in natural way the graph ${\cal G}(\xi_q)$ isomorphic to $G(\d_k)$.

\v 
{\bf Lemma 6.2}. {\it The number $F^{(q)}_1(3) = F_q\, c^{1-q}$ is given by the sum 
$$
F_q =cF_{q-1}+  \sum_{r =2}^q F_q(r),
\eqno (6.10)
$$
where the numbers $F_q(r), r \ge 2$  are  determined by the following recurrent relations
$$
F_q(r) =c \sum_{v =2}^r \ \sum_{u=0}^{q-v}
\ \sum_{s=0}^{q-v-u} 
{{ [{v-1\over 2}] + [{s\over 2}]}\choose{[{v-1\over 2}]}}
{ { [{r\over 2}]-1}\choose {[{v\over 2}] -1} } F_{q-u-v}(s) F_{u}(r-v).
\eqno (6.11)
$$
In (6.11),  $[x]$ denotes the largest integer less or equal to $x$ and  the following initial 
conditions are assumed:  $F_l(0) = \delta_{l,0}$, $F_{r-j}(r)=0$ for $j>0$ and 
$F_2(2)=1$. }

\v
{\it Remark.} Relations (6.10) and (6.11) are similar to those derived in \cite{BG,KV} for the number of walks 
of even number of steps such that their graph is a tree. 

\v {\it Proof.} Let us consider the set $\Xi_q(r)$ of  walks  such that there are $r$ steps 
that start or end in
$\rho$.
We  denote by $f_q(r) = \vert \Xi_q(r)\vert$ the cardinality of  this set. 

\begin{figure}[htbp]
\centerline{\includegraphics[width=12cm]{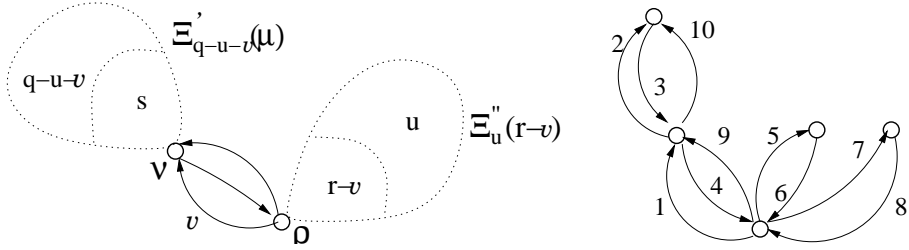}}
\caption{\footnotesize{The first-passage decomposition and example of the walk }}
\end{figure}

Let us consider the subset  $\Xi_q(r,v,s)\subset \Xi_q(r)$ consisting of the walks 
that pass  the edge $e_1= (\rho,\nu)$ $v$  times and that have $s$ steps attached to $\nu$ that do not pass $e_1$ (see Figure 7). 
We denote the set of such steps  $\nu\to\pi$ with $\pi\neq \rho$ by $\S_{\nu\to\bar \rho}$. Certainly, such a walk has
$r-v$ steps that belong to the set $\S_{\rho\to\bar \nu}$.

Any walk of $\Xi_q(r,v,s)$ can be  separated into three parts or in other words into the following three subwalks:
the first one contains the steps $\rho\to\nu$ and $\nu\to\rho$ only,
the second one is attached to $\nu$ with the elements of $\S_{\nu\to\bar \rho}$ (the left part of the walk) and the third one
is attached to $\rho$ by the elements of $\S_{\rho\to \bar \nu}$ (the right part of the walk). We denote the left and the right 
parts by $\Xi'_{q-u-v}(s)$ and by $\Xi''_u(r-v)$, respectively,
where $u$ denotes the number of steps in the right-hand part of the walk. 
We denote the by  $\Xi_q(r,v,s,u)$  the set of elements of  $\Xi_q(r,v,s)$ 
such that the right-hand part of the walk consists of $u$ steps.

Then we can write equality
$$
\vert \Xi_q(r,v,s,u)\vert  = {{ [{v-1\over 2}] + [{s\over 2}]}\choose{[{v-1\over 2}]}}
{ { [{r\over 2}]-1}\choose {[{v\over 2}] -1} }\, f_{q-u-v}(s) f_{u}(r-v),
\eqno (6.12)
$$
where we have taken into account obvious equalities 
$
 \vert \Xi'_{q-u-v}(s)\vert= \vert \Xi_{q-u-v}(s)\vert$ 
and  
$ 
 \vert \Xi''_u(r-v)\vert= \vert \Xi_u(r-v)\vert.
$

Let us explain the from of the combinatorial coefficients in  the right-hand side of (6.12).
 If $r$ and $v $ are both even, then
there are $v/2$ steps in $\S_{\nu\to\rho}$ and $r/2 - v/2$ steps in 
$\S_{\rho\to\bar\pi}$.
Therefore there is a choice to perform each of the $r/2 - v/2$ steps after one of $v/2$
steps $\nu\to\rho$. This produces the second binomial coefficient of (6.12). 

If $r$ is odd and $v$ is even, then the last step of the form $\rho\to\pi$ with $\pi\neq\rho$
is to be performed after that all the steps $\nu\to\rho$ are done. 
Then we get again the combinatorial coefficient as that given by (6.12). If $v$ is odd, then there are still $[{v/2}]$ 
steps $\nu\to\rho$.

Let us explain the first binomial coefficient of (6.12). If $s$ and $v$ are both even, 
then there are ${v/2 + s/2 -1\choose v/2-1}$ possibilities to perform $s/2$ steps
of $\S_{\nu\to\bar  \rho}$ after $v/2$ steps of $\S_{\rho\to\nu}$. If $s$ is odd, then the last step
is performed after that all $v/2$ steps are performed. If $v$ is odd, then there are $[v/2]+1$
steps $\rho\to\nu$.

Summing (6.12) with respect to parameters $\a, u$, and $s$, 
and taking into account that $\E a = c/n$, we get relation  (6.11) 
for the contribution to $F_q$ determined by the walks with $r, r\ge 2$ steps attached to $\rho$. 
If $r=1$, then corresponding contribution is given by $cF_{q-1}$. Summing all contributions, we obtain formula (6.10). Lemma is proved.  
\v
{\it Remark.}  It is easy to see from (6.12) that $f_3(2)=0$. Moreover, it is not hard to show that
$f_{2l+1}(2s) =0$ for all $l,s\ge 1$. This is in agreement with the obvious observation
that the set of walks $\Xi_{2l+1}(2,s)$ is empty.

\subsection {Cumulants of dilute and sparse $X$-model}

Analysis of cumulants of $X$-model in the dilute random graphs regime is very similar to that
of the $Y$-model in the same regime. 
\v
{\bf Lemma 6.4} {\it Relations (4.11) and (4.12) are  true with 
$$
\Phi_{k}^{(2q'+1)} (B) = (4q'+2)^{k-1} \quad {\hbox{and}} \quad \Phi_k^{(2q')} (B) =\left[ 
{1\over q'+1}{2q' \choose q'}\right]^{k} 2^{k-1} d^{(q')}_k,
\eqno (6.13)
$$
where $d^{(q')}_k$ is determined by relations (5.1) and (5.2) with $q$ replaced by $q'$.
}
\v 
{\it Proof.} 
According to Theorem 4.2, the leading contribution to (4.7) in the case of odd $q=2q'+1$  is given by the diagrams
$\d_k$ such that the graph $G(\d_k)$ is of maximal number of edges and contains one cycle only. 
Taking into account that each element $\l_j$ is represented by a cyclic graph with $2q'+1$ edges, 
the only possibility to get rid of the "extra" cycles in $\d_k$ and is to glue the elements $\l_j$ 
all long one of each other. To get maximal number of edges in $\d_k$, we keep the cyclic structure of $\l_j$.
When doing this, we have to choose the direction and the edge of $\l_1$ to glue the first elements of $\l_j, j=2, \dots, k$. 
The number of possibilities is given by $2^{k-1}\times (2q'+1)^{k-1}$ and we get the first equality of (6.13).

Regarding the case of even $q=2q'$, we  conclude that to construct a tree with the maximal number of edges we have first to construct a tree from each element $\l_j$ and then to draw $k-1$ arcs between these $k$ trees obtained. 
It is easy to see that the maximal size of the tree obtained by gluing the edges of $\l_j$ is 
$q'$ and the number of different trees is given by Catalan number 
$(q'+1)^{-1} {2q'\choose q'}$ \cite{S}.
Then we produce the connected diagrams exactly as in the case of dilute  $Y$ model. This gives the second relation of (6.13).
Lemma is proved.
\vskip 0.5 cm

Similarly to the  $Y$-model, the  sparse random graphs regime of $X$-model is 
more complicated than the dilute random graphs regime. The average value of random variable $X_n^{(2q')} = \T A^{2q'}$ is studied in \cite{BG,KV} (see also \cite{KSV}).
The limit $\Phi_1^{(2q')}(3) = \lim_{n\to\infty} {1\over nc^{q'}} \E X_n$ is determined
by a system of recurrent relations that have the form similar to that described in Lemma 6.2. We do not present these results.

The variance of $X_n^{(2q')}$ is studied  in \cite{V} in  more general setting than that of this paper.
The resulting expression is also determined by a system of recurrent relations.
This system  is much more  complicated that that for the first moment of $X_n$. We refer the reader to  the paper \cite{V} for corresponding theorems.

\subsection{Formal limit of the partition function}

Returning to the normalized partition function (2.12) that describes the $Y$-model with $q=2$, 
one can see that 
$$
{1\over p_nn^2} \log \hat Z_{n}(\b,tg_n) =  
{n-1\over 2p_n n}\log \left( {1 + e^{-2\b'}\over 1+e^{-2\b}}\right) +
{1\over p_nn^2} \log \E_{\b'}\left\{e^{tg_n Y_n}\right\},
\eqno (6.15)
$$
where $\b' = \b - tg_n$ and $g_n = (p_nn)^{-1}$. It is clear that 
the three asymptotic regimes introduced  in Section 4
are characterized by the corresponding behavior of the inverse "temperature" $\beta$ and 
elementary analysis shows that 
$$
\lim_{n\to\infty} {n-1\over 2p_n n}\log \left( {1 + e^{-2\b'}\over 1+e^{-2\b}}\right) =
\cases{ 0,& if $\beta = Const$;\cr
0, & if $\beta = {1\over 2} \log {n\over c},\ 1\ll c\ll n$; \cr
{\exp\{2tc^{-1}\} -1\over  2}, & if $\beta = {1\over 2} \log {n\over c},\ c= Const$,\cr}
$$
where $p_n = e^{-2\b}(1+e^{-2\b})^{-1}$. 
Then, assuming that the limit of the last term of (6.15) exists, we conclude that (4.4) implies relation
$$
\lim_{n\to\infty} {1\over p_nn^2} \log \hat Z_{n}(\b,tg_n) = {\d_{\omega,3}\over 2}  
\left(\exp\left\{{2t\over c}\right\} -1\right)
+ \sum_{k=1}^\infty {t^k F^{(2)}_k(\omega)\over k!},
\eqno (6.16)
$$
where $\omega $ indicates the full, the dilute, and the sparse random graphs regimes, and 
$\d_{\omega,3}$ denotes  the Kronecker $\d$-symbol.

Comparing expressions (6.1), (6.2), and (6.4) for $F_k^{(2)}$ with $\omega=1,2,3$, we see the difference between
the limiting rate functions of (6.16). Indeed, $F^{(2)}_k$ does not depend on $c$ in dilute random graphs regime, 
$\omega =2$, and is given by a polynomial in degrees of $1/c$ in the sparse random graph regime, $\omega =3$.
In particular, regarding the small-$t$ expansion of the right hand-side of (6.16) in the sparse random graphs limit
and taking into account that $F^{(2)}_1(3) = 1+c^{-1}$,  
we see that this approximation to the rate function is 
$
t(1 +{2\over c})(1+o(1)), t\to 0$.
The same difference between the rate functions is observed  in the general case of $q$, with exponential term of (6.16) replaced by 
$
{1\over 2}(\exp\{{2t\over c^{q-1}}\}-1)$.

Let us stress that the above reasoning relies strongly on the existence of the limit 
$$
\lim_{n\to\infty} {1\over p_n n^2} \log \E_{\b'} \left\{e^{tg_n V_n} \right\} = f_q(t)
\eqno (6.18)
$$
with $V_n  = X_n^{(q)}$ or $Y^{(q)}_n$. This problem is difficult to solve by using the cumulant expansion (3.1). As it is pointed out many times (see for example \cite{Ey}), the terms of the formal relation (3.1) with positive $g$ are used to enumerate combinatorial structures related to the matrix integrals. In the rigorous sense, the series (3.1) never converges in this case.
Existence and analyticity in $g$ of the terms of asymptotic expansion (3.1) with respect to $n$ is proved for the matrix integrals just recently \cite{BI,EM} by using powerful techniques of the integrable models and the Riemann-Hilbert problem.

\section{Summary }

The Gibbs weight 
determined by  the graph Laplacian
generates   a measure $\mu_n$ on the ensemble of $n$-dimensional adjacency matrices of simple non-oriented graphs.  This measure is invariant with respect to the permutations of the basis vectors and determines 
the Erd\H os-R\'enyi ensemble of random graphs with the edge probability $p_n$.

Regarding the sum over the set of weighted adjacency matrices as the analog of the matrix integrals,
we determine the discrete analog of matrix models with the quartic potential. In the general case of $q$-power potential, we distinguish two different families of discrete Erd\H os-R\'enyi matrix models. These are related with the numbers of $q$-step walks and $q$-step closed walks over the  random graphs denoted by $Y_n^{(q)}$ and $X_n^{(q)}$, respectively.

The logarithm of the standard partition function of these models is determined by the cumulants of random variables $X_n^{(q)}$ and $Y_n^{(q)}$. We develop a diagram technique to study the limiting behavior of the cumulants in three major asymptotic regimes of full, dilute and sparse random graphs 
determined by the edge probability $p_n$ as  $n\to\infty$.  We prove that the limits of these cumulants, when properly normalized, exist in all of the three asymptotic regimes. As a consequence, the Central Limit Theorem is shown to be true for centered and normalized variables $X$ and $Y$.  This implies CLT for the moments of the normalize spectral measure of the adjacency matrix of random 
Erd\H os-R\'enyi graphs.

We show that the limiting expressions of the cumulants are related with the number of non-rooted trees constructed with the help of labeled edges. In the simplest case of the dilute random graphs regime, the exponential generating function of these numbers $H(z)$ verifies the Lagrange (or P\'olya)  equation. 
Passing to the sparse random graphs regime, 
we derive more general equations that determine $H(z)$ in this case.

These results imply an observation that the asymptotic regimes we consider are different not only with  respect to the normalization factors of the cumulants, but also with the respect to the rate functions of large deviations formally determined as the limit 
$
f_q(t)
$ (6.18).
It should be noted that we did not prove rigorously the existence of this limit because our results concern the leading terms  of the cumulants only.

\end{document}